# MAJORS II: HCO$^+$ & HCN Abundances in W40

René Plume[1],[⋆] David J. Eden[2], Malcolm J. Currie[3], Lawrence K. Morgan[4] Xue-Jian Jiang[5], James DiFrancesco[6,7], Masatoshi Imanishi[8,9], Kee-Tae Kim[10,11], Tie Liu[12], Raffaele Rani[13], Alessio Traficante[14], Jane Cohen[1], Neal J. Evans II[15], Luis C. Ho[16,17], Eun Jung Chung[18], Sihan Jiao[19], Chang Won Lee[10,11], Dana Alina[20], Toby Moore[21], Jonathan M.C. Rawlings[22] Florian Kirchschlager[23], Sudeshna Patra[24], Andrew J. Rigby[25], Hsien Shang[26], Jihye Hwang[27], Patricio Sanhueza[28], Mark G. Rawlings[29], Kianoosh Tahani[30], Junfeng Wang[31], Kate Pattle[22], James S. Urquhart[32], Quang Nguyen-Luong[33], Sarah E. Ragan[34], Yang Su[35], Xindi Tang[36], Agata Karska[37,24], and Michael G. Burton[38]

[1] Dept. of Physics & Astronomy, University of Calgary, 2500 University Dr. NW, Calgary, AB, Canada, T2N 1N4
[2] Department of Physics, University of Bath, Claverton Down, Bath BA2 7AY, UK
[3] STFC RAL Space, Rutherford Appleton Laboratory, Harwell Campus, Didcot, Oxfordshire OX11 0QX, UK.
[4] Green Bank Observatory, 155 Observatory Rd, Green Bank, WV 24944, USA
[5] Research Centre for Astronomical Computing, Zhejiang Laboratory, Hangzhou 311121, P.R.China
[6] NRC Herzberg Astronomy and Astrophysics, 5071 West Saanich Rd, Victoria, BC V9E 2E7, Canada
[7] Department of Physics and Astronomy, University of Victoria, Victoria, BC V8W 2Y2, Canada
[8] National Astronomical Observatory of Japan, Osawa, Mitaka, Tokyo 181-8588, Japan
[9] Department of Astronomical Science, School of Science, Graduate University for Advanced Studies (SOKENDAI), Mitaka, Tokyo 181-8588, Japan
[10] Korea Astronomy and Space Science Institute, 776 Daedeokdae-ro, Yuseong-gu, Daejeon 34055, Republic of Korea
[11] University of Science and Technology, Korea (UST), 217 Gajeong-ro, Yuseong-gu, Daejeon 34113, Republic of Korea
[12] Shanghai Astronomical Observatory: Shanghai, CN
[13] Institute of Astronomy, National Tsing Hua University, No. 101, Section 2, Kuang-Fu Road, Hsinchu 30013, Taiwan
[14] IAPS – INAF, Via Fosso del Cavaliere, 100, I-00133 Rome, Italy
[15] Department of Astronomy, The University of Texas at Austin, 2515 Speedway, Stop C1400 Austin, Texas 78712-1205, USA
[16] Kavli Institute for Astronomy and Astrophysics, Peking University, Beijing 100871, China
[17] Department of Astronomy, School of Physics, Peking University, Beijing 100871, China
[18] Korea Astronomy and Space Science Institute, 776 Daedeokdae-ro, Yuseong-gu, Daejeon 34055, Republic of Korea
[19] National Astronomical Observatories, Chinese Academy of Sciences, 20A Datun Road, Chaoyang District, Beijing 100012, China
[20] Physics Department, Nazarbayev University, Astana, Kazakhstan
[21] Astrophysics Research Institute, Liverpool John Moores University, IC2, Liverpool Science Park, 146 Brownlow Hill, Liverpool L3 5RF, UK
[22] Department of Physics & Astronomy, University College London, Gower Street, London WC1E 6BT, UK.
[23] Sterrenkundig Observatorium, Ghent University, Krijgslaan 281-S9, B9000 Gent, Belgium
[24] Institute for Advanced Studies, Nicolaus Copernicus University in Toruń, Wileńska 4, 87-100 Toruń, Poland
[25] School of Physics and Astronomy, University of Leeds, Leeds LS2 9JT, UK
[26] Institute of Astronomy and Astrophysics, Academia Sinica: Taipei, Taiwan, TW
[27] Korea Astronomy and Space Science Institute (KASI), 776 Daedeokdae-ro, Yuseong-gu, Daejeon 34055, Republic of Korea
[28] Department of Astronomy, School of Science, The University of Tokyo, 7-3-1 Hongo, Bunkyo, Tokyo 113-0033, Japan
[29] Gemini Observatory/NSF NOIRLab, 670 N. A'ohōkū Place, Hilo, HI 96720, USA
[30] Department of Physics & Astronomy, Kwantlen Polytechnic University, 12666 72nd Avenue, Surrey BC V3W 2M8, Canada
[31] Department of Astronomy, Xiamen University, Xiamen, Fujian 361005, China
[32] Centre for Astrophysics and Planetary Science, University of Kent, Canterbury, CT2 7NH, UK
[33] Department of Computer Science, Mathematics & Environmental Science, The American University of Paris, PL111, 2 bis, passage Landrieu 75007 Paris, France
[34] Cardiff Hub for Astrophysics Research & Technology, School of Physics & Astronomy, Cardiff University, Queen's Buildings, Cardiff CF24 3AA
[35] Purple Mountain Observatory, Yuanhua Road 10, Nanjing 210023, China
[36] Xinjiang Astronomical Observatory, CAS 150, Science 1-Street, 830011 Urumqi, PR China
[37] Max-Planck-Institut für Radioastronomie, Auf dem Hügel 69, 53121, Bonn, Germany
[38] Armagh Observatory and Planetarium, College Hill, Armagh, Northern Ireland, BT61 9DB, UK

**ABSTRACT**

We present observations of HCN and HCO$^+$ J = 3 → 2 in the central $424'' \times 424''$ region of the W40 massive star forming region. The observations were taken as part of a pilot project for the MAJORS large program at the JCMT telescope. By incorporating prior knowledge of N(H$_2$) and $T_K$, assuming a constant density, and using the RADEX radiative transfer code we found that the HCN and HCO$^+$ abundances range from $X$(HCN) = $0.4 - 7.0 \times 10^{-8}$ and $X$(HCO$^+$) = $0.4 - 7.3 \times 10^{-9}$. Additional modelling using the NAUTILUS chemical evolution code, that takes H$_2$ density variations into account, however, suggests the HCN and HCO$^+$ abundances may be fairly constant. Careful modelling of three different positions finds $X$(HCN) = $1.3 - 1.7 \times 10^{-8}$, $X$(HCO$^+$) = $1.3 - 3.1 \times 10^{-9}$. Cross-comparison of the two models also provides a crude estimate of the gas density producing the HCN and HCO$^+$ emission, with H$_2$ densities in the range $5 \times 10^4 - 5 \times 10^5$ cm$^{-3}$ – suggesting that the HCN and HCO$^+$ emission does indeed arise from dense gas. High UV intensity (e.g. $G_o >$ a few thousand) has no effect on the abundances in regions where the visual extinction is large enough to effectively shield the gas from the UV field. In regions where $A_V < 6$, however, the abundance of both species is lowered due to destructive reactions with species that are directly affected by the radiation field.

## 1 INTRODUCTION

Stars form in the densest parts of molecular clouds (Lada et al. 2010) and, in high-mass star forming regions, the densities can reach $10^6$ cm$^{-3}$ or higher (Plume et al. 1997). Typical molecular tracers like CO and $^{13}$CO J = 1 → 0 measure the total H$_2$ gas mass in molecular clouds reasonably well. However, since the low J transitions of CO can be easily excited in gas with densities < 100 cm$^{-3}$, they primarily trace the low density ambient molecular gas in Giant Molecular Clouds (GMCs), not specifically the gas that will contribute to forming stars. In addition, CO is known to freeze out onto dust grains in high-density, low-temperature environments (e.g. Thomas & Fuller 2008; Fontani et al. 2012; Sabatini et al. 2019) and may, therefore, be an unreliable tracer of the gas under such conditions.

Higher critical density molecular line tracers, such as HCN, probe the behaviour of the dense gas most closely associated with star formation very well (Onus et al. 2018). A survey of spiral galaxies and Ultra Luminous Infrared Galaxies (ULIRGS) in HCN J = 1 → 0 revealed a tight relationship between the infrared luminosity ( L$_{IR}$), a tracer of the Star Formation Rate (SFR), and the HCN luminosity (L$_{HCN}$), a tracer of total dense molecular gas content (Gao & Solomon 2004a; Gao & Solomon 2004b; Stephens et al. 2016). When these results are extended down to nearby star-forming regions, i.e. Galactic clumps, the linear relationship survives, covering a luminosity range many orders of magnitude wide and scales from parsec-sized clumps to entire ULIRGS (Wu et al. 2005; Mendigutía et al. 2018). There are, however, some notable exceptions. For example, Kauffmann et al. (2017) and Santa-Maria et al. (2023) showed that HCN $J$ = 1 → 0 traces densities of < $10^4$ cm$^{-3}$ in cold regions of Orion A and Orion B, respectively. Furthermore, Evans et al. (2020) demonstrated that this transition can also arise in diffuse gas (n < 100 cm$^{-3}$) provided the column density is sufficiently high (A$_V$ > 8).

While simulations have shown that the L$_{IR}$ - L$_{HCN}$ relation is a result of the free-fall timescale of the gas and the average gas density of molecular clouds (Krumholz et al. 2012), both Galactic and extragalactic observations have pointed towards the presence of a volume-density threshold (> $10^4$ cm$^{-3}$), with the amount of gas above this density being the important factor in regulating the SFR (Lada et al. 2012; Evans et al. 2014; Zhang et al. 2014; Heiderman et al. 2010). It is, therefore, clear that dense gas is the important fuel for star formation. Understanding its exact role, and how the properties of molecular clouds impact the process, however, has been elusive.

The Massive, Active, JCMT-Observed Regions of Star formation (MAJORS) survey is a large program at the 15m James Clerk Maxwell Telescope (JCMT) on Maunakea, Hawaii. MAJORS was awarded 976 hours of band 4 weather conditions (i.e. 4-6 mm of precipitable water vapour) to simultaneously map HCO$^+$ J = 3 → 2 ($\nu$ = 267.5576259 GHz) and HCN J = 3 → 2 ($\nu$ = 265.8864339 GHz) in a large mass-selected sample of dust-continuum traced molecular clouds in the Milky Way. The sample includes clouds in the Central Molecular Zone (CMZ), the Inner Galaxy, and the Outer Galaxy. While full details of the MAJORS survey will be presented in Eden et al. (2026, in prep), its primary goals are to:

(i) understand how Galactic environment influences the physics of dense gas, enabling us to determine how dense gas is produced and how it is intrinsically linked to star formation;

(ii) produce L$_{IR}$–L$_{gas}$ relationships for dense gas in Galactic clumps and molecular clouds, allowing us to connect these results to studies of extragalactic systems and ULIRGs and to assess the universality of the star-formation process;

(iii) determine the origin of variations in the HCN and HCO$^+$ abundances and HCN/HCO$^+$ abundance ratio, and how these variations relate to the physical conditions imposed by different Galactic environments.

HCN and HCO$^+$ J = 3 → 2 are excellent tracers of dense gas. In the optically thin limit, the critical densities of HCO$^+$ and HCN J = 3 → 2 are $1.4 \times 10^6$ cm$^{-3}$ and $1.0 \times 10^7$ cm$^{-3}$ respectively (Shirley 2015). However, since these lines are often optically thick, radiative trapping effectively reduces their critical densities to $6.8 \times 10^3$ cm$^{-3}$ and $7.3 \times 10^4$ cm$^{-3}$ respectively (Shirley 2015) in which case these lines may not be tracing the innermost, densest regions of the clouds.

* E-mail: rplume@ucalgary.ca

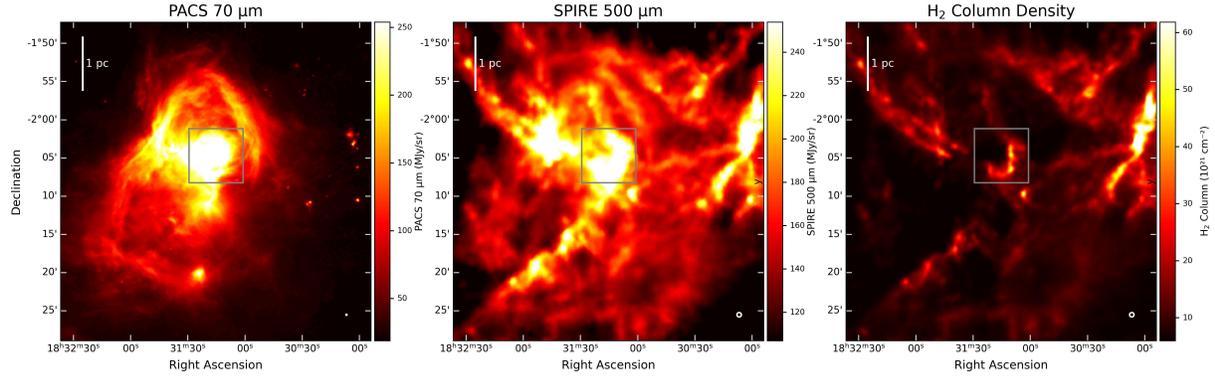

**Figure 1.** (Left) Public PACS-70 $\mu$m observations and (middle) SPIRE-500 $\mu$m observations from the Herschel Science Archive. (Right) H$_2$ column density maps from the Herschel Gould Belt Survey (HGBS; André et al. 2010; Bontemps et al. 2010; Könyves et al. 2015; Arzoumanian et al. 2019) (see Section 2). The green box shows the region that we focus on in this paper (see e.g. Figure 2). A 1 parsec scalebar is shown at the upper left corner of each image, and a circle indicating the observational beamsize appears in the lower right.

| Object(s) | $X$(HCN) | $X$(HCO$^+$) | $X$(HCN)/$X$(HCO$^+$) | Resolution | Reference |
|---|---|---|---|---|---|
| Low-Mass Protostars | $1.7 \times 10^{-9}$ | $6.6 \times 10^{-9}$ | 2.6 | 13-18″ | Jørgensen et al. (2004) |
| W51 | $2.0 \times 10^{-9}$ | $1.5 \times 10^{-10}$ | 13.3 | 1800″ (40-50 pc) | Watanabe et al. (2017) |
| YSO GL2591 | $2.0 \times 10^{-8}$ | $10^{-8}$ | 2.0 | 13″ | van der Tak et al. (1999) |
| IRDCs - Quiescent | – | $2.4 \times 10^{-8}$ | – | 38″ | Sanhueza et al. (2012) |
| IRDCs - Active | – | $5.7 \times 10^{-8}$ | – | 38″ | Sanhueza et al. (2012) |
| ATLASGAL Clumps - Quiescent | – | $4.0 \times 10^{-9}$ | – | 38″ | Hoq et al. (2013) |
| ATLASGAL Clumps - Protostellar | – | $6.4 \times 10^{-9}$ | – | 38″ | Hoq et al. (2013) |
| ATLASGAL Clumps - HII/PDR | – | $7.5 \times 10^{-9}$ | – | 38″ | Hoq et al. (2013) |
| 70$\mu$m Quiet Clumps | – | $2.5 \times 10^{-10}$ | – | 27″ | Traficante et al. (2017) |
| Starburst Galaxies | – | – | 2–7 | 50 pc | Butterworth et al. (2025) |
| AGN | – | – | 1–41 | 50 pc | Butterworth et al. (2025) |
| ULIRGS | – | – | 2–24 | 500 pc | Butterworth et al. (2025) |
| LIRGS & ULIRGS | $10^{-7} - 10^{-8}$ | $10^{-8}$ | 1–10 | 0.3″ (50-500 pc) | Nishimura et al. (2024) |

**Table 1.** HCN and HCO$^+$ abundances (relative to N(H$_2$)) for various objects in the literature.

In either case, however, these transitions can probe gas with densities associated with star formation activity rather than the low density bulk envelopes of molecular clouds.

Given the high critical densities of these transitions, many previous studies have simply correlated total HCN or HCO$^+$ luminosity with total infrared luminosity to measure the star formation rate in dense gas (see e.g. Stephens et al. (2016) and references therein). However this interpretation is overly simplistic. *In order to be certain that a particular transition of these molecules is actually tracing gas above the $10^4$ cm$^{-3}$ threshold, one must know their abundances*; given by the quantity $X$, where $X$ is calculated by dividing the molecular column density by the column density of H$_2$ (e.g. $X$(HCN) = N(HCN)/N(H$_2$) and $X$(HCO$^+$) = N(HCO$^+$)/N(H$_2$)).

The abundances of HCN and HCO$^+$ (and other molecular species) are not constant, however, and can vary from location to location. Table 1 lists HCN and HCO$^+$ abundances (and their ratios) for a range of Galactic and extragalactic sources observed at different resolutions. These values span nearly two orders of magnitude between sources. Some of this variation may reflect differences in beam size or in the methods used to derive molecular abundances; however, Table 1 clearly demonstrates that intrinsic variations exist. In addition, beyond observational and methodological effects, molecular abundances are fundamentally shaped by environmental factors such as gas and dust temperature, ultraviolet radiation fields, opacity, and cosmic-ray ionization rate. To fully understand the connection between dense gas, $L_{\rm IR}$, and star formation in the Galaxy, it is therefore essential to measure HCN and HCO$^+$ abundances (and associated gas densities) across a large, diverse sample of sources spanning many Galactic environments, using a consistent methodology to minimize systematic uncertainties.

### 1.1 W40

As part of a pilot project for MAJORS, we mapped HCO$^+$ and HCN J = 3 → 2 in the central region of the well-known massive star forming region W40. Recent astrometry observations from Gaia (Comerón et al. 2022) place W40 at a distance of 502 ± 4 pc, which makes it one of the nearest high-mass star forming regions; comparable in distance to the Orion Molecular Cloud. W40 lies on the far side of the Aquila Rift and so, in the visible, it is partially obscured by foreground extinction (up to visual extinctions of A$_V$ ~ 3 magnitudes; Straižys et al. 2003). Thus, it is best observed at Infrared (IR) and Far Infrared (FIR) wavelengths. Figure 1 shows a 70$\mu$m (left) and a 500$\mu$m (middle) image of the W40 complex using the PACS (Poglitsch et al. 2010) and SPIRE (Griffin et al. 2010) instru-ments (respectively) aboard the ESA Herschel Space Observatory[1] (Pilbratt et al. 2010). The data were obtained from the Herschel Science Archive[2]. Figure 1 (right) shows the H$_2$ column density maps derived by the Herschel Gould Belt Survey (HGBS) (André et al. 2010; Bontemps et al. 2010; Könyves et al. 2015; Arzoumanian et al. 2019) (see Section 2). The 500$\mu m$ (Figure 1: middle) and H$_2$ column density (Figure 1: right) images shows that W40 is associated with an extended molecular cloud complex with an estimated mass of ∼ $10^4$ M$_\odot$ (Rodney & Reipurth 2008). The 70$\mu m$ image shows two cavities that form an hour-glass (bipolar) shape on large scales (a few pc) with an HII region at the centre powered by an OB association. The bright-rimmed clouds at the cavity walls show clearly that dense clumps and pillars are illuminated from inside by the cluster. The main cluster is located just northwest of the intersection of the two cavities seen in Figure 1 (left); near the tip of the "hook" within the grey/green box in the H$_2$ column density map (Figure 1 - right). The cluster is more clearly seen in the zoomed-in image of the 2MASS H-band data (Skrutskie et al. 2006) in Figure 2 (top right). The OB association is comprised of IRS/OS1a (O9.5), IRS/OS2b (B4) and IRS/OS3a (B3) and an associated stellar cluster of pre-main-sequence stars (Shuping et al. 2012). Chandra X-ray observations (Kuhn et al. 2010) have revealed over 200 young stellar objects in a $17' \times 17'$ area centred on the cluster. Based on the Herschel fluxes at 70 and 160 $\mu m$, the average strength of the UV field in W40 is 237 $G_o$ (Schneider et al. 2020; and references therein) where $G_o$ is the strength of the average interstellar radiation field in Habing units ($1.6 \times 10^{-3}$ erg cm$^{-2}$ s$^{-1}$; Habing 1968). However, the same observations show that the UV field strength can reach ∼ 8200 $G_o$ near the OB association.

In this paper we present observations and analysis of HCN and HCO$^+$ J = 3 → 2 in W40, taken as part of the MAJORS survey at the JCMT. The main goals of this paper are to determine the HCN and HCO$^+$ abundances, examine the variations in abundance and physical conditions across the entire region mapped, and to estimate the gas density associated with the HCN and HCO$^+$ emission. In Section 2 we present the observations. In Section 3 we present basic analyses of the data including integrated intensity maps and the observational relationship between the HCN and HCO$^+$ integrated intensities. In Section 4 we calculate the HCN and HCO$^+$ abundances based on an analysis of the data utilizing both the RADEX 1D radiative transfer code (van der Tak et al. 1999) and the NAUTILUS gas-grain chemical evolution code (Ruaud et al. 2016). In Section 5 we present the main conclusions of the paper.

## 2 OBSERVATIONS

The HCO$^+$ J = 3 → 2 line at $\nu$ = 267.5576259 GHz and HCN J = 3 → 2 line at $\nu$ = 265.8864339 GHz were observed as part of the MAJORS large program (Eden et al. 2026; in prep - observing program M22AL002) at the James Clerk Maxwell Telescope (JCMT). Line frequencies were obtained from the Cologne Database for Molecular Spectroscopy (CDMS)[3] (Müller et al. 2001; Müller et al. 2005; Endres et al. 2016). The data were taken using the dual polarization, dual-sideband ʼŬʼŭ heterodyne receiver (Kerr et al. 2014). Maps were centred at $\alpha$(J2000) = $18^h$ $31^m$ $17^s$ and $\delta$(J2000) = $-02°$ $05'$ $00''$. Full Width at Half Maximum (FWHM) beam sizes are $\theta_{HCO^+\ J=3\to2}$ = 18.3″ and $\theta_{HCN\ J=3\to2}$ = 18.4″. Data were obtained on 2020 August 21 as part of director's discretionary time (observing program M20AD003) to test the MAJORS observing strategy.

The atmospheric opacity at 225 GHz ranged from 0.225 to 0.315 over the course of the observations (as measured by the JCMT's in-cabin 183 GHz Water Vapour Monitor). Data were obtained in raster scan, position switching mode with a scan speed of 16″/sec and an off position at $\alpha$(J2000) = $18^h$ $31^m$ $53.7^s$ and $\delta$(J2000) = $-1°$ $59'$ $35''$. Two scans in orthogonal directions (i.e. "basket-weaved") were used to produce a smoother output map. Maps are 424″ × 424″ in size with a pixel/sampling size of 8″ × 8″. The median system temperature during the observations was ∼300 K. The spectrometer, ACSIS, was used in the 250 MHz mode (8192 channels, 30.517578 kHz spacing) resulting in a native spectral resolution of 0.034 km s$^{-1}$.

Data reduction was performed with ORAC-DR (Jenness et al. 2015), which is built on the STARLINK (Currie et al. 2014) packages KAPPA (Currie & Berry 2014), CUPID (Berry et al. 2007), and SMURF (Chapin et al. 2013). We used the REDUCE_SCIENCE_NARROWLINE recipe that creates a spatial cube from the raw time-series data, removes a baseline from the data, and trims the ends of the frequency range to remove high-noise regions. The data cubes provided by the pipeline were subsequently converted from NDF to FITS format using the STARLINK NDF2FITS command to allow for further processing in CASA (McMullin et al. 2007) and PYTHON. Post-processing, the data were spectrally rebinned in STARLINK by 15 channels to provide a spectral resolution of ∼ 0.5 km s$^{-1}$. With this resolution, the average 1-$\sigma$ noise is 0.33 K for the HCO$^+$ J = 3 → 2 data and 0.29 K for the HCN J = 3 → 2 data ($T_A^*$). This noise was estimated using a robust baseline-subtracted MAD method with Savitzky–Golay smoothing of 3 km s$^{-1}$, polyorder = 2, $\sigma$-clipping at 2.5$\sigma$, and 3 iterations.

The HCO$^+$ and HCN data in this paper are presented in the corrected antenna temperature ($T_A^*$) scale, but are converted to main beam brightness ($T_{mb} = T_A^*/\eta_{mb}$) for column density calculations (Section 4.1) using a main beam efficiency of $\eta_{mb}$ = 0.66[4]. While calibration errors for the ʼŬʼŭ receiver are not fully characterized, the JCMT receiver webpage[5] states that "typical uncertainties in peak flux measurements tend to be in the range of 10-20%". Based on this, we adopted an average calibration uncertainty of 15%.

We note that HCN has hyperfine structure (e.g. Mullins et al. 2016; Loughnane et al. 2012) but for the J = 3 → 2 transition, the inner four lines are only 0.35, 0.07, and –0.54 km s$^{-1}$ from the central strongest line and, given the width of our observed lines (> 1 km s$^{-1}$; see Section 3), are blended, appear as a single component, and cannot be distinguished from one another. The outer two lines are 1.82 and –2.28 km s$^{-1}$ from the central, strongest line but, since these are each 25 times weaker than the central component (and they are not detected), we ignore the hyperfine structure of HCN in the remainder of this paper.

For subsequent analysis, we also obtained dust/kinetic temperature and H$_2$ column density maps from the Herschel Gould Belt Survey (André et al. 2010; Bontemps et al. 2010; Könyves et al. 2015; Arzoumanian et al. 2019; hereafter HGBS). These maps were produced by Könyves et al. (2015) by taking the Herschel SPIRE data at 250, 350, and 500 μm, smoothing to the resolution of the SPIRE 500 μm observations ($\theta_{\text{FWHM}} = 36.3''$) and fitting the data to a modified blackbody of the form: $I_\nu = B_\nu(T_d) \kappa_\nu \mu_{H_2} m_H N(H_2)$, where $\kappa_\nu$ is the dust mass opacity coefficient (given as $\kappa_\lambda = 0.1(\lambda/300\mu m)^{-2}$ cm$^2$/g), $\mu_{H_2}$ is the mean molecular weight per hydrogen molecule (2.8), and $m_H$ is the mass of atomic hydrogen. In the fit, the dust temperature ($T_d$) and column density (N(H$_2$)) were left as free parameters. The resulting images were sampled on a 3'' pixel grid. Details of the procedure to produce the dust/kinetic temperature and H$_2$ column density maps are provided in Könyves et al. (2015). These maps were subsequently regridded to the same grid and pixel size as our HCO$^+$ and HCN data (i.e. 8'' × 8'') using the CASA (McMullin et al. 2007) `imregrid` command.

---

[1] Herschel was an ESA space observatory with science instruments provided by European-led Principal Investigator consortia and with important participation from NASA.
[2] http://archives.esac.esa.int/hsa/whsa/
[3] https://cdms.astro.uni-koeln.de/classic/
[4] https://www.eaobservatory.org/jcmt/instrumentation/heterodyne/namakanui/uu-230ghz/
[5] https://www.eaobservatory.org/jcmt/instrumentation/heterodyne/calibration/

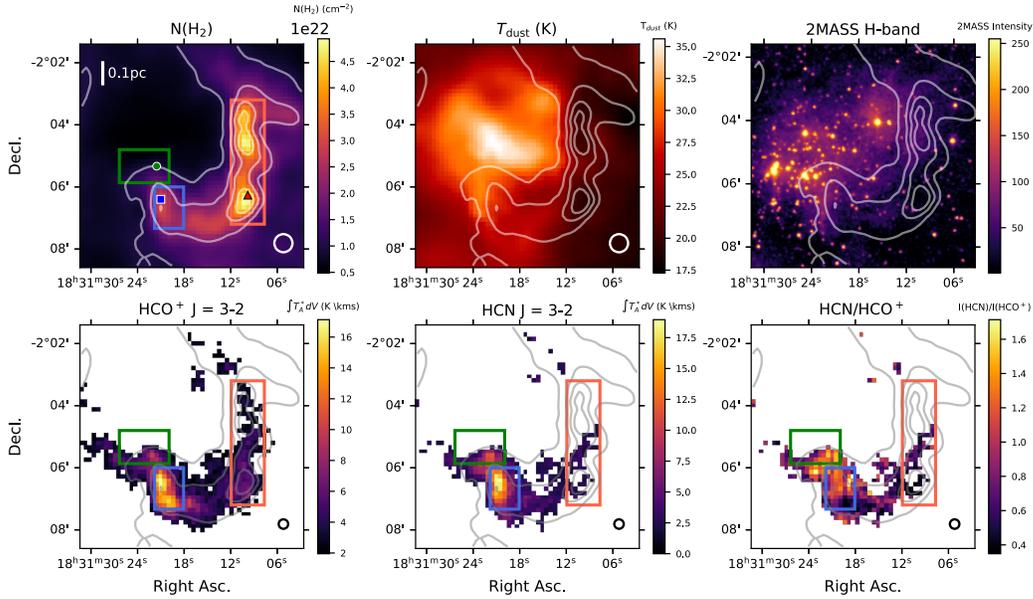

**Figure 2.** (Top left) HGBS H$_2$ column density overlaid with contour levels of 1,2,3,4 ×10$^{22}$ cm$^{-2}$. These same contours are overlaid in all following plots. The green, blue, and red boxes indicate regions selected for low H$_2$ column density & high temperature, intermediate N(H$_2$) & $T_K$, and high N(H$_2$) & low $T_K$ respectively. The Green Circle, Blue Square, and Red Triangle denote positions for which we perform chemical modelling in Section 4.3. (Top middle) HGBS dust temperature. (Top right) 2MASS H-band image. (Bottom left) Integrated intensity image for HCO$^+$ J = 3 → 2. (Bottom middle) Integrated intensity image for HCN J = 3 → 2. (Bottom right) Ratio of the HCN/HCO$^+$ integrated intensity. Values for each quantity are given in the colour bars to the right of each plot. A 0.1 parsec scalebar is shown at the upper left corner of the first image, and a circle indicating the observational beamsize appears in the lower right of each image (except for the 2-MASS image).

## 3 RESULTS

### 3.1 Integrated intensity comparisons

Figure 2 shows the H$_2$ column density (N(H$_2$) - top left) and dust/kinetic temperature ($T_{\text{dust}} = T_K$ - top middle) from the HGBS, as well as the location of young stars seen in the 2MASS H-band image (top right) (Skrutskie et al. 2006). Note that we will continue to use the dust temperature as a proxy for the gas kinetic temperature, but will explore the effects of this assumption on our analysis in Section 4.1.2.

The figure also shows the integrated intensity maps for HCO$^+$ J = 3 → 2 (bottom left) and HCN J = 3 → 2 (bottom middle), and the ratio of the HCN/HCO$^+$ integrated maps (bottom right).

To derive the HCO$^+$ and HCN integrated intensities, we fit a single Gaussian to every pixel in the HCO$^+$ and HCN cubes using the Trust Region Reflective Least Squares fitter (TRFLSQFitter) in Python/Astropy. This fitter allows upper and lower parameter bounds and, in Astropy 5.3 and later, provides covariance-based parameter uncertainties. The fits yield peak temperatures ($T_A^*$; K), line widths ($\Delta V_{\text{FWHM}}$; km s$^{-1}$, with $\Delta V_{\text{FWHM}} = 2\sqrt{2\ln 2}\sigma$), line centroids ($V_{\text{LSR}}$; km s$^{-1}$), and integrated intensities ($\int T_A^* \, dV$; K, km s$^{-1}$, with $\int T_A^* \, dV = T_A^* \sigma \sqrt{2\pi}$). Errors in the integrated intensity were calculated from the errors in $T_A^*$ and $\sigma$ from the equation $\sqrt{(\sigma \times \sigma_{\text{err}})^2 + (T_A^* \times T_{A,\text{err}}^*)^2} \times \sqrt{2\pi}$, where $T_{A,\text{err}}^*$ is calculated by combining the 15% calibration error in quadrature with the Gaussian fitting error. We retain only fits with Signal-to-Noise (S/N) > 3 in *both* the peak and the integrated intensity; pixels failing either criterion are set to *NaN*. Accepted fits were visually inspected to reject spurious solutions. We adopt Gaussian-fit integrals rather than MOMENT maps because this enables simultaneous S/N cuts on peak and area; tests with the MOMENT command, however, produced consistent results. Blank regions in the integrated-intensity maps mark pixels with S/N < 3. Per-pixel uncertainties depend on the local noise but are, on average, ∼25 per cent of the integrated intensity.

The green, blue, and red boxes in Figure 2 delineate regions chosen for their locations relative to the OB association and for specific conditions. Figure 3 shows that points within the Green Region generally have higher kinetic temperatures (< $T_K$ > = 30.2 K) and lower H$_2$ column densities (< N(H$_2$) > = 1.26 × 10$^{22}$ cm$^{-2}$) than the rest of W40. In contrast, points within the Red Region have some of the lowest kinetic temperatures (< $T_K$ > = 21.6 K) and highest H$_2$ column densities (< N(H$_2$) > = 3.5 × 10$^{22}$ cm$^{-2}$). Points in the Blue Region have intermediate mean values of $T_K$ and N(H$_2$) of 27 K and 2.4 × 10$^{22}$ cm$^{-2}$ respectively. These regions will be utilized more fully in later discussion (see Section 4). Figure 4 shows example spectra and Gaussian fits at the positions marked by the Green Circle, Blue Square, and Red Triangle in Figure 2 (top left).

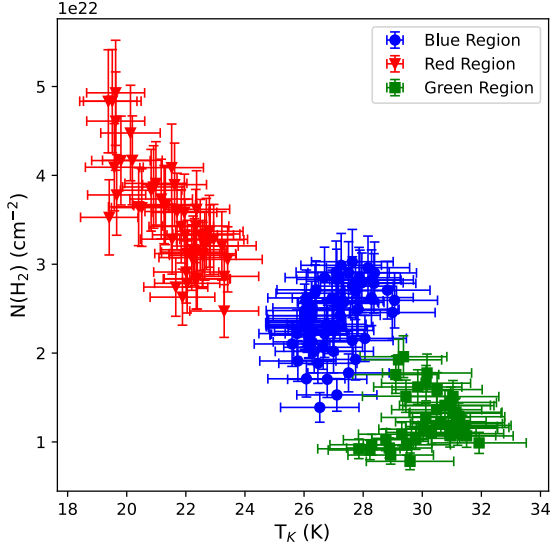

**Figure 3.** N(H$_2$) versus $T_K$ (i.e. $T_{dust}$) for every pixel within the coloured boxes shown in Figure 2. The green points plot the values for pixels located within the Green Region, the blue points plot the values for pixels located within the Blue Region, and the red points plot the values for pixels located within the Red Region. Error bars are 5 per cent in $T_K$ and 12 per cent in N(H$_2$) based on the analysis of uncertainties by Roy et al. 2014.

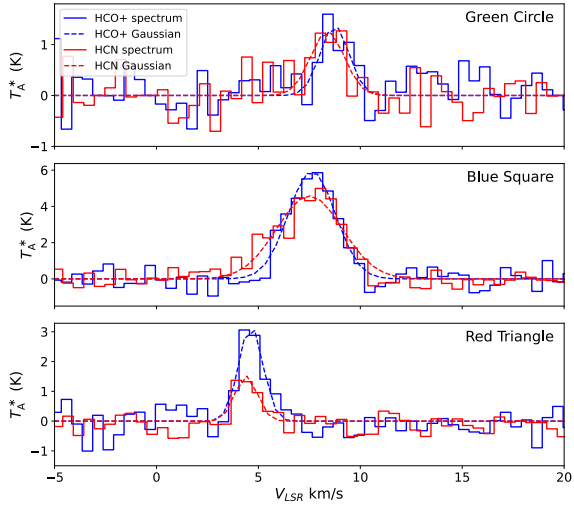

**Figure 4.** HCN and HCO$^+$ spectra and their Gaussian fits of three representative positions marked by the Green Circle, Blue Square, and Red Triangle in Figure 2.

These spectra also indicate that there is a velocity gradient across the filament, which first moment maps reveal to be quite smooth.

As can be seen in Figure 2, the molecular gas is distributed in a curved filament wrapping around the hottest gas which is co-located with the central OB association. While the HCO$^+$ and HCN have similar morphologies and extent, the HCO$^+$ is slightly more extended than HCN in the low integrated intensity regions. This is particularly noticeable to the left of the Green Region (where the H$_2$ column density is low and the dust temperature is high) and in the Red Region (where the H$_2$ column density is high but the dust temperature is low). The lower number of HCN pixels in the Red Region is due to weaker HCN emission lines (see e.g. Figure 4 - bottom) that causes the integrated intensity to fall below the 3-$\sigma$ noise limit.

Figure 5 shows the relationship between the HCN J = 3 → 2 and HCO$^+$ J = 3 → 2 integrated intensity ($I$(HCN) and ($I$(HCO$^+$) respectively) for the points displayed in Figure 2. Using the stats.linregress function within SCIPY we found that, for the entire data set (i.e. points both inside and outside the Red, Green, and Blue Regions in Figure 2), $I$(HCN) (K km s$^{-1}$) = 0.96 $I$(HCO$^+$) (K km s$^{-1}$) with a y-intercept of –0.70, a Pearson r-value of 0.83, and a p-value of < 10$^{-9}$. This suggests that there is a strong linear correlation to the data and that the trend is statistically significant. Over the entire region <$I$(HCN)> = 4.9 K and <$I$(HCO$^+$)> = 4.8 K.

Breaking the data down into the regions displayed in Figure 2, the coloured points show the relationship for each of the respective coloured regions, with solid black lines showing the best linear fit. In the Blue Region (left panel) $I$(HCN) = 0.89 $I$(HCO$^+$) with a y-intercept of 0.23, a Pearson r-value = 0.75, and a p-value of < 10$^{-9}$; again suggesting strong evidence for a statistically significant linear relationship. In this region <$I$(HCN)> = 9.6 K and <$I$(HCO$^+$)> = 10.5 K. In the Red Region (middle panel) $I$(HCN) = 0.14 $I$(HCO$^+$) with a y-intercept of 1.93, a Pearson r-value = 0.28, and a p-value of 0.05; suggesting that there is only weak evidence for a statistically significant, linear relationship. In this region <$I$(HCN)> = 2.6 K and <$I$(HCO$^+$)> = 4.8 K. And, in the Green Region (green points) $I$(HCN) = 1.48 $I$(HCO$^+$) with a y-intercept of –2.26, a Pearson r-value of 0.81, and a p-value of < 10$^{-9}$. In this region <$I$(HCN)> = 6.6 K and <$I$(HCO$^+$)> = 5.9 K. Thus, the correlation is weakest in the Red Region, where the H$_2$ column density is highest, the dust temperature is the lowest, and the line intensities are weakest. The Green and Blue Regions, on the other hand, have similar, and strongly correlated, relationship between the HCN and HCO$^+$ integrated intensities.

The relationship between our J = 3 → 2 integrated intensities in the Blue Region is similar to that seen in the J = 1 → 0 transition in outer Galaxy clouds, where $I$(HCN J = 1 → 0) = 0.5 to 0.9 $I$(HCO$^+$ J = 1 → 0) (Patra et al. 2022; Braine et al. 2023). In the inner Galaxy, a sample of 6 clouds by Evans et al. (2020) find $I$(HCN) J = 1 → 0) = 1.2 $I$(HCO$^+$ J = 1 → 0); more similar to the relationship seen in the warm, Green Region. Given the lower critical densities of the 1 → 0 transitions and the larger beamsize of these observations (∼ 1′), it is surprising to see such a similarity in these relationships. It should be noted, however, that in the case of Evans et al. (2020) and Patra et al. (2022), the authors did attempt to restrict their analysis to dense gas by only selecting high column density/extinction regions with $A_V$ > 8.

The fit to the full dataset closely resembles that derived for the Blue Region. This is likely because the Blue Region spans the widest range of integrated intensities and contains the brightest HCN and HCO$^+$ emission. As a result, the global fit is strongly influenced by the large number of high-intensity points originating in this region. This illustrates the risk of examining the HCN–HCO⁺ relationship in a purely global sense, as bright regions can mask distinct relationships that may be present in other parts of the same cloud.

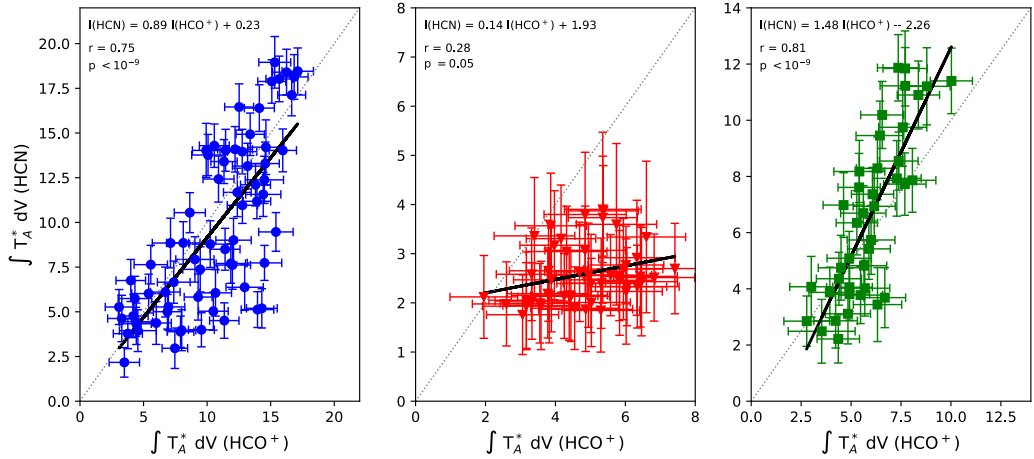

**Figure 5.** Integrated intensities (K km s$^{-1}$) of HCN J = 3 → 2 versus HCO⁺ J = 3 → 2 for all positions with a signal to noise ratio above 3 − $\sigma$ (i.e. those displayed in Figure 3 and Figure 2). Points are colour coded to indicate the blue (left panel), red (middle panel), or green (right panel) regions (seen in Figure 2) from which they were taken. The solid black lines showing the best linear fit to the data points in each of the regions. The dotted grey line shows shows the $I$(HCN) = $I$(HCO⁺) relation for comparison.

### 3.2 Molecular line ratios

Over the entire map, the HCN/HCO⁺ integrated intensity ratio ranges from 0.35 to 1.71 with a mean value of 0.83 ($\sigma$ = 0.3). As can be seen in Figure 2, while the HCN/HCO⁺ ratio traces the general morphology of the HCN integrated intensity map, it does not peak at the same location as the HCN or HCO⁺ emission peaks. In fact, the HCN/HCO⁺ ratio is moderately larger (average of $\approx$ 1.1) to the north of the regions of brightest HCN and HCO⁺ emission ( i.e. in the Green Region and the northern part of the Blue Region; towards the hot gas surrounding the OB association) than it is throughout the bulk of the filament (where the average is $\approx$ 0.5 to 0.6). In addition, points within the Green Region generally have elevated HCN/HCO⁺ integrated intensity ratios (< $I$(HCN)/$I$(HCO⁺) > = 1.1). In contrast, the Red Region encompasses some of the lowest HCN/HCO⁺ integrated intensity ratios (< $I$(HCN)/$I$(HCO⁺) > = 0.5). Points in the Blue Region have an intermediate value of < $I$(HCN)/$I$(HCO⁺) > = 0.9.

These kinds of differences in the HCN/HCO⁺ ratio are also seen in the observations of Nguyen-Luong et al. (2020) who mapped the HCN and HCO⁺ J = 1 → 0 emission in M17 and found that the mean HCN/HCO⁺ value is enhanced in M17-HII (1.36) versus that in M17-IRDC (1.09); a difference that they attribute to increased UV heating in M17-HII. We can make more direct comparisons, however, with HCN/HCO⁺ J = 4 → 3 observations due to the similar critical densities between the J = 3 → 2 and 4 → 3 transitions. In the Perseus cloud Walker-Smith et al. (2014) find a HCN/HCO⁺ J = 4 → 3 ratio of 0.1 to 0.6; which is similar to what is seen in the bulk of the filament in W40. On larger (i.e. Galactic) scales, Tan et al. (2018) found that the HCN/HCO⁺ J = 4 → 3 ratio in six nearby star forming galaxies varies from 0.1 to 2.7 with a mean value of 0.9 (standard deviation of 0.6), which is very comparable to our observed values across the whole of the W40 region. Similarly, Butterworth et al. (2025) studied the HCN/HCO⁺ J = 3 → 2 and J = 4 → 3 ratios in a sample of 80 sources (active galactic nuclei, AGN; starburst galaxies, SB; and (U)LIRGs) at various spatial resolutions. At molecular-cloud scales (25–50 pc), the HCN/HCO⁺ J = 3 → 2 ratio is < 1 in SB galaxies, consistent with our red and Blue Regions. In the Green Regions, the ratio exceeds unity but does not reach typical AGN values (> 1.5). This suggests that, while similar physical/chemical processes may influence the line ratios in the Green Regions, the environmental conditions are less extreme than in AGN.

## 4 DISCUSSION

### 4.1 HCO⁺ & HCN column densities

To determine the HCO⁺ and HCN column densities, we utilized the Gaussian fits to the spectra as described in Section 3. The peak and integrated intensities of the valid fits were converted from $T_A^*$ to $T_{mb}$ units using the main beam efficiency (see Section 2). Each pixel with an accepted Gaussian fit was then assigned a gas kinetic temperature ($T_K$) from the regridded HGBS dust temperature map (Figure 2) assuming that $T_K = T_{dust}$. We will explore the effects of this assumption in Section 4.1.2.

We then utilized the RADEX 1D radiative transfer code (van der Tak et al. 2007) which, given a density, kinetic temperature, line width, and molecular column density, produces the expected peak intensity ($T_{mb}$) and opacity for a particular molecular transition. The HGBS maps allowed us to fix the temperature in each pixel. The Gaussian fitting also provided the observed integrated intensity ($\int T_{mb} dV$) in each pixel. We then used the standard Python script

| | | $X$(HCN) ($\times 10^{-8}$) | $X$(HCO$^+$) ($\times 10^{-9}$) | $X$(HCN)/$X$(HCO$^+$) |
|---|---|---|---|---|
| Entire Map | | | | |
| | Mean | 2.0±0.5 | 1.9±0.5 | 9.1±0.3 |
| | Min | 0.4±0.2 | 0.4±0.2 | 2.6±0.7 |
| | Max | 7.0±1.0 | 7.3±1.1 | 23.1±9.0 |
| Green Region | | | | |
| | Mean | 3.5±0.7 | 3.0±0.7 | 12.0±0.6 |
| | Min | 1.3±0.5 | 1.2±0.5 | 5.3±1.7 |
| | Max | 6.0±0.9 | 5.6±1.1 | 19.1±6.0 |
| Blue Region | | | | |
| | Mean | 3.4±0.6 | 4.0±0.7 | 9.5±0.4 |
| | Min | 1.0±0.3 | 1.0±0.4 | 2.6±0.7 |
| | Max | 7.0±1.0 | 7.3±1.1 | 23.1±9.0 |
| Red Region | | | | |
| | Mean | 0.7±0.2 | 1.3±0.3 | 6.1±0.4 |
| | Min | 0.4±0.2 | 0.4±0.2 | 3.1±1.1 |
| | Max | 1.1±0.3 | 2.7±0.5 | 14.8±8.0 |

**Table 2.** HCN, HCO$^+$ abundances (relative to N(H$_2$)) and HCN/HCO$^+$ abundance ratios for various regions within W40. Results are presented for our fiducial case with n(H$_2$) = 10$^5$ cm$^{-3}$.

RADEX-COLUMN.PY (available on the RADEX website[6]), which begins with an initial column density of $10^{12}$ cm$^{-2}$ and iteratively adjusts the molecular column density (up or down) until the RADEX model integrated intensity matches the observed value within a specified tolerance. Thus, we were able to determine the HCO$^+$ and HCN column densities for every pixel in our HCO$^+$ and HCN maps with accepted Gaussian fits. The only free parameter in our RADEX calculation was the (unknown) H$_2$ **volume** density (n(H$_2$)), so we sampled densities logarithmically from $10^3$ to $10^8$ cm$^{-3}$ in 0.5-dex increments (i.e., $10^3$, $5 \times 10^3$, $10^4$, $5 \times 10^4$, ...). Accordingly, we generated HCN and HCO$^+$ column density maps for each of the volume densities. The optimal density is determined later through comparison with the predictions of the chemical models, which are described in Section 4.3.

### 4.1.1 Effects of assuming local thermodynamic equilibrium (LTE)

Although LTE column densities are easier to calculate, LTE assumes that the population of every level is well described by the Boltzmann distribution; an assumption which is certainly not appropriate for the densities explored here. To illustrate, we have compared an optically thin LTE analysis to a "fiducial" RADEX result which uses a volume density of n(H$_2$) = 10$^5$ cm$^{-3}$. We have chosen this H$_2$ density to represent a fiducial result because it is midway between the density extremes of $10^3 - 10^8$ cm$^{-3}$ used in our analysis in the previous Section.

The comparison between the optically thin LTE analysis and the fiducial RADEX results shows a large difference between the two. The RADEX HCN column densities are 50-100 times larger than the LTE values and the RADEX HCO$^+$ column densities are 5-20 times larger than the LTE values. These differences are due to both excitation and opacity effects. At a density of n(H$_2$) = 10$^5$ cm$^{-3}$, the rotational levels are sub-thermally excited and so the Boltzmann equation does not properly account for the population in the other states. In addition, the RADEX derived opacity (for n(H$_2$) = 10$^5$ cm$^{-3}$) across the entire cloud is significant; with $\tau_{HCN} \sim 5-80$ and $\tau_{HCO^+} \sim 1-20$. Thus, at a density of n(H$_2$) = 10$^5$ cm$^{-3}$ the optically thin, LTE assumption is seriously in error, and one must take both opacity and non-LTE excitation into account.

[6] https://sronpersonalpages.nl/ vdtak/radex/index.shtml

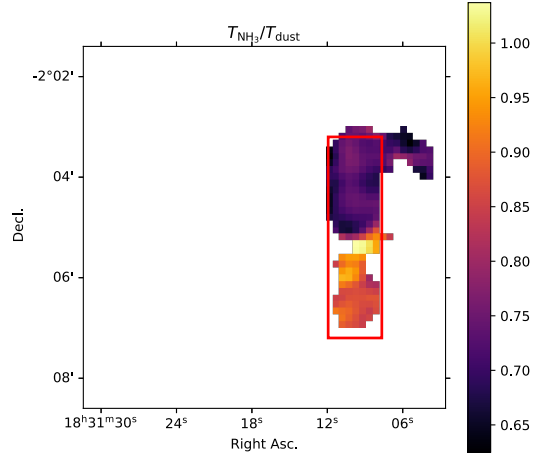

**Figure 6.** Map of the ratio of the kinetic temperature calculated from NH$_3$ observations (Friesen et al. 2017; $T_{NH_3}$) to the dust temperature calculated from the HGBS data (Könyves et al. 2015; $T_{dust}$). The red rectangle denotes the same "Red Region" as indicated in Figure 2

### 4.1.2 Effects of assuming $T_K = T_{dust}$

In the previous analysis we assumed that the HCN and HCO$^+$ kinetic temperature was provided by the dust temperature maps (i.e. $T_K = T_{dust}$). This assumption is probably sufficient in regions where n(H$_2$) > $10^5$ cm$^{-3}$ (e.g. Goldsmith 2001; Kruegel & Walmsley 1984; Goldreich & Kwan 1974) but, in lower density regions, this assumption breaks down due to insufficient coupling between the gas and dust.

To test this assumption, and the affect it has on our column density calculations, we utilized the results of Friesen et al. (2017), who mapped the NH$_3$ (1,1) and (2,2) transitions in a number of sources using the Green Bank Telescope. By modelling the NH$_3$ data, they were able to produce maps of the gas kinetic temperature (that we call $T_{NH_3}$). Sensitivity in the NH$_3$ observations have limited this temperature map to the Red Region of our observations. These data were also regridded onto the same grid and pixel size as the HCO$^+$, HCN, and dust data using the CASA (McMullin et al. 2007) `imregrid` command.

Figure 6 shows the ratio of the gas to dust temperature (i.e. $T_{NH_3}$ / $T_{dust}$) in the Red Region of W40. There is a distinct difference in this ratio between the northern and southern sections of the Red Region. In the northern section, the average value of $T_{NH_3}$ / $T_{dust}$ is 0.71 ($\sigma$ = 0.03) whereas in the southern section the average value is 0.89 ($\sigma$ = 0.05). The reason for this dichotomy is not immediately clear but it may be related to the volume density. If the volume density is lower in the northern section, which could be true given the weak HCN and HCO$^+$ lines in this area, then we would expect weaker coupling between the gas and dust and, therefore, a poorer match between the gas and dust temperatures.

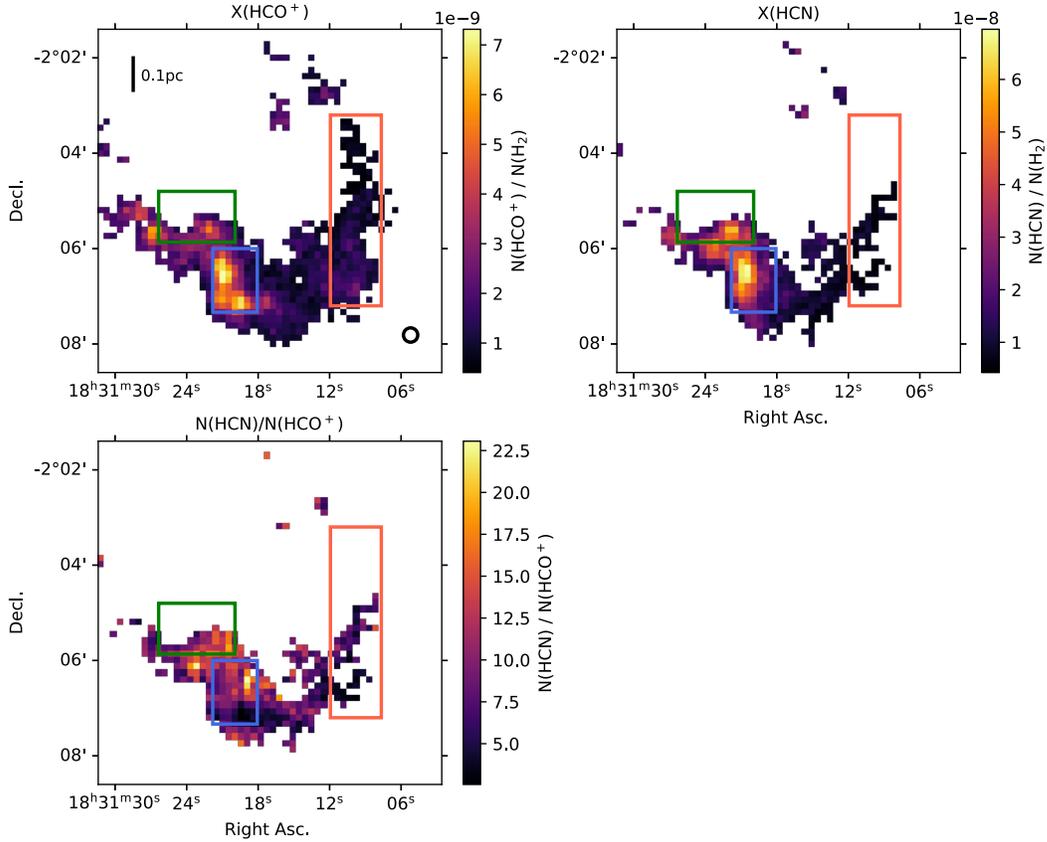

**Figure 7.** Abundance maps for (top left) HCO⁺ and (top right) HCN. The bottom left image shows the HCN/HCO⁺ column density (i.e. abundance) ratio. All values were calculated from the RADEX derived column densities using n(H$_2$) = 10$^5$ (for illustrative purposes), divided by the HGBS H$_2$ column density map. Colour bars to the right of each figure provide the relevant scales. A 0.1 parsec scale bar is shown in the upper left corner of the first image, and a circle indicating the observational beam size appears in the lower right.

The difference in temperature corresponds to a difference in the derived HCN and HCO⁺ column densities as well. Using our fiducial volume density (i.e. n(H$_2$) = 10$^5$ cm$^{-3}$) we again used RADEX to calculate the HCN and HCO⁺ column densities using both the dust temperature map (N($T_{\rm dust}$)) and the NH$_3$ temperature map (N($T_{\rm NH_3}$)). A map of N($T_{\rm NH_3}$) / N($T_{\rm dust}$) reveals a similar distinct ratio between the northern and southern sections of the Red Region. For HCN, in the northern section the average value of N($T_{\rm NH_3}$) / N($T_{\rm dust}$) is 1.78 ($\sigma$ = 0.13) whereas, in the southern section, the average value is 1.18 ($\sigma$ = 0.08). For HCO⁺, in the northern section the average value of N($T_{\rm NH_3}$) / N($T_{\rm dust}$) is 1.68 ($\sigma$ = 0.17) whereas, in the southern section, the average value is 1.22 ($\sigma$ = 0.10). For both molecules, the lower NH$_3$ temperature in the northern section results in larger column densities.

Therefore, in regions where the gas and dust temperatures are fairly well matched (i.e within 10 per cent), the HCN and HCO⁺ column densities are also fairly well matched. In this case, using the assumption that $T_{\rm K}$ = $T_{\rm dust}$, we underestimate column densities by only ∼ 20 per cent. However, in regions where the gas and dust temperatures are less well matched (i.e within 30 per cent), the HCN and HCO⁺ column densities can be underestimated by up to a factor of 2. The HCN/ HCO⁺ column density **ratio**, however, is not as strongly affected, since changes in the temperature affect both column densities by about the same amount and in the same direction.

### 4.2 HCN and HCO⁺ abundances

Our final step was to produce abundance maps by dividing the HCO⁺ and HCN column density maps produced by RADEX, by the regridded HGBS H$_2$ column density map (Figure 2). The HCO⁺ and HCN abundance maps ($X$(HCO⁺) and $X$(HCN) respectively) are shown in Figure 7 for our fiducial case with n(H$_2$) = 10$^5$ cm$^{-3}$, along with the ratio of the HCN to HCO⁺ abundance ($X$(HCN)/$X$(HCO⁺)). Note that these results are meant to reflect the abundance variations that one obtains by assuming a constant volume density across the entire region; an assumption that is almost certainly not accurate and will be explored in Section 4.3. Statistics for Figure 7 are provided in Table 2.

To estimate the abundance uncertainties, we propagated each pixel's integrated-intensity uncertainty (∼25 per cent) through RADEX. From the fitted integrated intensity we computed a fiducial column density, and then a maximum (minimum) column density by increasing (decreasing) the intensity by its uncertainty. We also tested a 5 per cent uncertainty in the kinetic temperature. Its effect on the derived column densities was negligible and is therefore ignored. From the resulting HCN and $HCO^+$ column densities and their uncertainties, we combined the 12 per cent $N(H_2)$ uncertainty (Roy et al. 2014) in quadrature to obtain fiducial HCN and $HCO^+$ abundances and their uncertainties. Although the propagated column density (and hence abundance) uncertainties are slightly asymmetric, at the precision reported in Table 2 they are effectively symmetric, so we quote them as simple ± values.

In Figure 7, we can see that the $HCO^+$ and HCN abundances mostly track with the integrated intensity maps (Figure 2) with peaks in the Blue Region (around $\alpha = 18^h\ 31^m\ 21^s$, $\delta = -2°\ 06'\ 30''$) and a "tail" extending into the Red Region (where $N(H_2)$ is high and $T_K$ is low). There are, however, a few notable differences. The Green Region and the northern part of the Blue Region (where $N(H_2)$ is low but $T_K$ is high) have enhanced $HCO^+$ and HCN abundances. Similar enhancements are not as clearly seen in the integrated intensity maps. In addition, the abundance ratio in these two regions is also enhanced, leading to $X(HCN)/X(HCO^+)$ ratios in excess of 12 (and up to ∼ 20), versus $<\sim 10$ in the rest of the cloud. Note that the region with $X(HCN)/X(HCO^+) > 15$ is also the region in which $I(HCN)/I(HCO^+) > 1$ (Section 3).

Table 2 highlights the regional trends discussed above. In particular, the Red Region has the lowest mean, minimum, and maximum values for both abundances and the abundance ratios (although the minimum abundance ratio is slightly larger than that seen in the Blue Region). The Blue Region has a mean HCN abundance similar to that of the Green Region but also has the highest mean $HCO^+$ abundance, resulting in a mean abundance ratio between that of the Red and Green Regions. While the absolute abundances vary by a factor of 15–20 across the map, the abundance ratio varies by only a factor of 9.

The column densities derived from RADEX and, therefore, the abundances, scale roughly with the density. If we decrease the density to $n(H_2) = 5 \times 10^4$ cm$^{-3}$ the abundances of both species scale upwards by a factor of ≈ 2 to 5, but the abundance ratios remain approximately constant (within ∼ ±20 per cent). Increasing the density to $n(H_2) = 5 \times 10^5$ cm$^{-3}$ results in a decrease in the absolute abundances by a factor of approximately 2 to 5, while the abundance ratios again remain largely unchanged. So, changes in the $H_2$ density can help explain the abundance variations across W40, as well as the relatively small changes in the abundance ratios (e.g. Table 2). A more detailed discussion of how density affects our results, as well as the method we use to estimate the appropriate $H_2$ volume density, is provided in the following Section.

### 4.3 Chemical modelling of the $HCO^+$ & HCN abundances

#### 4.3.1 Description of the modelling

To investigate the origin of the observed variations in HCN and $HCO^+$ abundances and abundance ratios, and to obtain a crude estimate of the gas density in the emitting regions, we modelled their time evolution with the NAUTILUS gas–grain chemical code (Ruaud et al. 2016). NAUTILUS follows chemical evolution in three phases – gas, grain surface, and grain mantle – and tracks how abundances change over time as a function of physical conditions such as: gas density, dust and gas temperature, visual extinction, ultraviolet flux, and cosmic-ray ionization (and others not discussed here). The code employs the KInetic Database for Astrochemistry (KIDA) database (Wakelam et al. 2012) and solves a series of coupled differential equations for a network of more than 10,000 reactions: 7,627 gas-phase reactions and 3,799 grain reactions, including adsorption, desorption, surface chemistry, mantle processing, and migration between the mantle and surface. These reactions connect 510 gas-phase species, 225 grain-surface species, and 225 grain-mantle species. No deuterated species or isotopologues are included in the reaction network.

As before, the HGBS maps allow us to fix the temperature and $H_2$ column density in each pixel. We have again assumed that $T_{dust} = T_{gas}$ everywhere. The $H_2$ column density provides us with the visual extinction in each pixel. Studies of the $H_2$ to $A_V$ conversion range from $A_V = \frac{N_{H_2}}{2.1 \times 10^{21}}$ (Rachford et al. 2009; Zhu et al. 2017) to $A_V = \frac{N_{H_2}}{1.9 \times 10^{21}}$ (Bohlin et al. 1978; Whittet 1981). In this paper, we have used the more recent value of $A_V = \frac{N_{H_2}}{2.1 \times 10^{21}}$.

For the UV field, the NAUTILUS code expresses the flux as $S \times 10^8$ photons cm$^{-2}$ s$^{-1}$ in the 6–13.6 eV range (Ruaud et al. 2016; Öberg et al. 2007), where $S$ is a scaling factor. Setting $S = 1$ corresponds to $G_0 \sim 1$ in Habing units, where $G_0 = 1.6 \times 10^{-3}$ erg cm$^{-2}$ s$^{-1}$ (Habing 1968). We used a scaling factor of $S = 237$ (i.e. $G_o = 237$) to model the average radiation field in W40 from its stellar cluster (Schneider et al. 2020). For the initial atomic abundances, we used the default values provided by Ruaud et al. (2016) (their Table 1) that assumes an oxygen abundance of $2.4 \times 10^{-4}$ and a C/O ratio of 0.7. We also used a standard cosmic-ray (CR) ionization rate of $\zeta_{H_2} = 1.3 \times 10^{-17}$ s$^{-1}$.

We employed a three-phase, static model within NAUTILUS. In this framework, the cloud is assumed to have fixed physical parameters that remain constant as the chemistry evolves over time. Although we acknowledge that real molecular clouds evolve dynamically, in the absence of information about the prior state and time evolution of the gas, we are constrained to adopt static conditions. The binding energies are taken from the KIDA database (Wakelam et al. 2012), with values of 2050 K for HCN and 1150 K for CO. No binding energy is provided for $HCO^+$, as NAUTILUS does not permit the freeze-out of ionic species such as $HCO^+$. Instead, interactions between ions and grains are treated through ion–grain recombination reactions, for example: $HCO^+ + grain^- \rightarrow CO + H + grain^0$, rather than through an adsorption–desorption balance.

The physical parameter with the most uncertainty is the $H_2$ volume density and so we ran NAUTILUS with densities of $n(H_2) = 10^3 - 10^8$ cm$^{-3}$ with the same logarithmic spacing as described in Section 4.1. Densities less than $10^4$ cm$^{-3}$ and greater than $10^6$ cm$^{-3}$ failed to match the abundances calculated by our RADEX analysis, so we restricted further analysis to $n(H_2) = 10^4, 5 \times 10^4, 10^5, 5 \times 10^5$, and $10^6$ cm$^{-3}$. We compared the model abundances from NAUTILUS calculated at a particular choice of $H_2$ density with those calculated from RADEX at the same $H_2$ density. If NAUTILUS was able to produce an HCN abundance, $HCO^+$ abundance, and HCN/$HCO^+$ abundance ratio that equaled those of RADEX then we considered a solution (or a "match") had been found. To decide which $H_2$ density provided the best solution, we performed this comparison for each $H_2$ density and chose the density for which the match between NAUTILUS and RADEX occurs at approximately the same time (or with the smallest time difference between them). To provide an additional constraint on the models, we required that the time at which the HCN/HCO⁺ abundance ratio matched the RADEX solution fall between the times when the individual HCN and HCO⁺ abundances matched the observations.

### 4.3.2 Results of the modelling

Figure 8 shows the modelling performed for three specific positions in W40 : the Green Circle, Blue Square, and Red Triangle in Figure 2. These positions were chosen since they have disparate physical conditions in which to investigate abundance variations. In particular, Figure 8 shows the time evolution of HCN (red), HCO⁺ (blue), and CO (black; scaled down by a factor of 1000) abundances, and the abundance ratio (orange). In all three cases, the time interval between when the HCN and HCO⁺ abundances match (i.e. the span of the vertical shaded region) is relatively small, and the time at which the HCN/HCO⁺ abundance ratio matches (vertical dotted line) falls within this interval, though near its outer limit. Note that while HCN crosses the observed abundance value twice, we adopt the second crossing time as the correct solution, since it also corresponds to the solution for the HCN/HCO⁺ ratio. Results for the specific positions are discussed individually below. Appendix A presents example model results across a range of densities and illustrates how specific density solutions were evaluated, showing why some were rejected and others accepted as the best match to the observed abundances.

<u>The Green Circle - The Low N(H₂) and High $T_K$ Regime:</u> The Green Circle represents a position with a low H₂ column density (N(H₂) = $9.9 \times 10^{21}$ cm⁻²; $A_V$ = 4.7) and high temperature ($T_K$ = 31.9 K). The best solution for this position occurs for an H₂ density of $10^5$ cm⁻³. At this density the RADEX derived abundances are $X$(HCN) = $1.72 \times 10^{-8}$ and $X$(HCO⁺) = $1.33 \times 10^{-9}$ which are close to average. The HCN/HCO⁺ abundance ratio, however, is relatively large (12.9). In this NAUTILUS model, both observed abundances and the abundance ratio are reached at ~ $4 - 6.5 \times 10^4$ years. It is important to emphasize that we are not interpreting this as the actual age of the cloud. As noted by Rawlings et al. (2024), chemical ages have limited meaning when modelling a dynamically evolving cloud under fixed physical conditions. Our goal is simply to demonstrate that if the HCN and HCO⁺ emission arise from the same gas under identical physical and dynamical conditions, then the precise evolutionary time is not critical. What matters is the ability to simultaneously reproduce the observed abundances of HCN and HCO⁺. The results for the best NAUTILUS model are given in the top panel of Figure 8.

<u>The Blue Square - The Moderate N(H₂) and $T_K$ Regime:</u> The Blue Square represents a position with a moderate H₂ column density (N(H₂) = $2.9 \times 10^{22}$ cm⁻²; $A_V$ = 13.8) and temperature ($T_K$ = 28.4 K). The best solution for this position occurs for an H₂ density of $5 \times 10^5$ cm⁻³. At this density the RADEX derived abundances are $X$(HCN) = $1.29 \times 10^{-8}$ and $X$(HCO⁺) = $1.29 \times 10^{-9}$ which are similar to those seen at the position of the Green Circle (although the Blue Square position has a higher H₂ density). The HCN/HCO⁺ abundance ratio, is also relatively large (10.0). In this NAUTILUS model, both observed abundances and the abundance ratio are reached at ~ $3.2 - 4.6 \times 10^4$ years. The best matching solution is shown in the middle panel of Figure 8.

<u>The Red Triangle - The High N(H₂) and Low $T_K$ Regime:</u> The Red Triangle represents a position with a high H₂ column density (N(H₂) = $4.6 \times 10^{22}$ cm⁻²; $A_V$ = 21.9) and a low temperature ($T_K$ = 19.6 K). It is also furthest from the stellar cluster. The best solution for this position occurs for an H₂ density of $5 \times 10^4$ cm⁻³. At this density the RADEX derived abundances are $X$(HCN) = $1.27 \times 10^{-8}$ and $X$(HCO⁺) = $3.13 \times 10^{-9}$. While the HCN abundance is similar to that seen at the positions of the Green Circle and the Blue Square, the HCO⁺ abundance here is a factor of ~ 3 higher resulting in a fairly low HCN/HCO⁺ abundance ratio (4.1). In this NAUTILUS model, both observed abundances and the abundance ratio are reached at ~ $1.5 - 2.4 \times 10^5$ years. The results for this NAUTILUS model are given in the bottom panel of Figure 8.

The results for all three positions are summarized in Table 3 that shows that, in W40, the HCN abundance ranges from $X$(HCN) = $1.27 - 1.72 \times 10^{-8}$, and HCO⁺ abundance varies from $X$(HCO⁺) = $1.29 - 3.13 \times 10^{-9}$ and the $X$(HCN)/$X$(HCO⁺) ratio varies from 4.1 – 12.9. Thus, allowing the density to vary from $5 \times 10^4 - 5 \times 10^5$ cm⁻³ results in smaller abundance variations than is seen when the density is held constant (see Figure 7 and Table 2).

Density variations of this magnitude are reasonable within molecular clouds. For example, a study of the CS $J = 1 \to 0$, $J = 2 \to 1$, and $J = 5 \to 4$ transitions of a number of selected clumps within Orion A found that the density varied from $10^4 - 2.2 \times 10^5$ cm⁻³ (Tatematsu et al. 1998). Also, using a simple assumption that the depth of the H₂ molecular cloud is equivalent to its width ($\approx 0.1$ pc - see Figure 2) the H₂ volume density in W40 would vary from $\approx 3 \times 10^4 - 1.5 \times 10^5$ cm⁻³. Under this simplifying assumption of a constant line-of-sight depth (0.1 pc), the Red Region, having the highest column density, should also exhibit the highest volume density, not the lowest as reported in Table 3. However, an alternative geometry cannot be ruled out; the Red Region may trace the edge of a limb-brightened shell wrapping around the stellar cluster rather than a simple column. In this scenario, low H₂ volume density could exist in a region of high column density.

### 4.3.3 Effects of Density on HCN and HCO⁺ Abundances and Line Strengths

The H₂ density affects the time-evolution of each species due to their different formation and destruction pathways. The primary formation route for HCO⁺ at this temperature is:

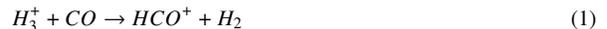
$$H_3^+ + CO \to HCO^+ + H_2 \quad (1)$$

and the main destruction pathway is:

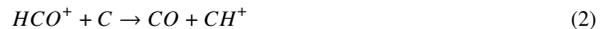
$$HCO^+ + C \to CO + CH^+ \quad (2)$$

Thus, for HCO⁺, the important factors are the amount of CO available to form HCO⁺, the amount of neutral atomic carbon available to destroy it, and the times at which these two forms dominate the carbon budget. For low densities (e.g. n(H₂) = $5 \times 10^4$ cm⁻³), neutral atomic C is the dominant form of carbon between $10^2$ and $3 \times 10^4$ years. However, at higher densities (e.g. n(H₂) = $5 \times 10^5$ cm⁻³), neutral atomic C is the dominant form of carbon at an earlier time, i.e. between 10 and $3 \times 10^3$ years. Therefore, at higher densities, the conversion from C to CO occurs very early in the simulation so one simultaneously loses the (destructive) neutral atomic carbon and increases the amount of (productive) CO at much earlier times than one does at low densities.

In addition, as shown in Figure 8, the blue position — with the highest density — reaches a large CO abundance earliest in the simulation (at ~ 6500 years, before it temporarily depletes), allowing the HCO⁺ abundance to build up quickly and match the observed value earliest in the simulations. The red position — with the lowest density — reaches its maximum CO abundance much later ($4.6 \times 10^4$ years), and therefore reproduces the observed HCO$^+$ abundance latest. As expected, the green position — with intermediate density — reaches the observed HCO$^+$ abundance at a time between the blue and red positions. The overall result is that the modelled HCO$^+$ abundance agrees with the observed value at earlier times in regions of higher-density gas. It is also noteworthy that the green and Blue Regions do not undergo the same CO depletion seen in the red position, likely due to their elevated temperatures.

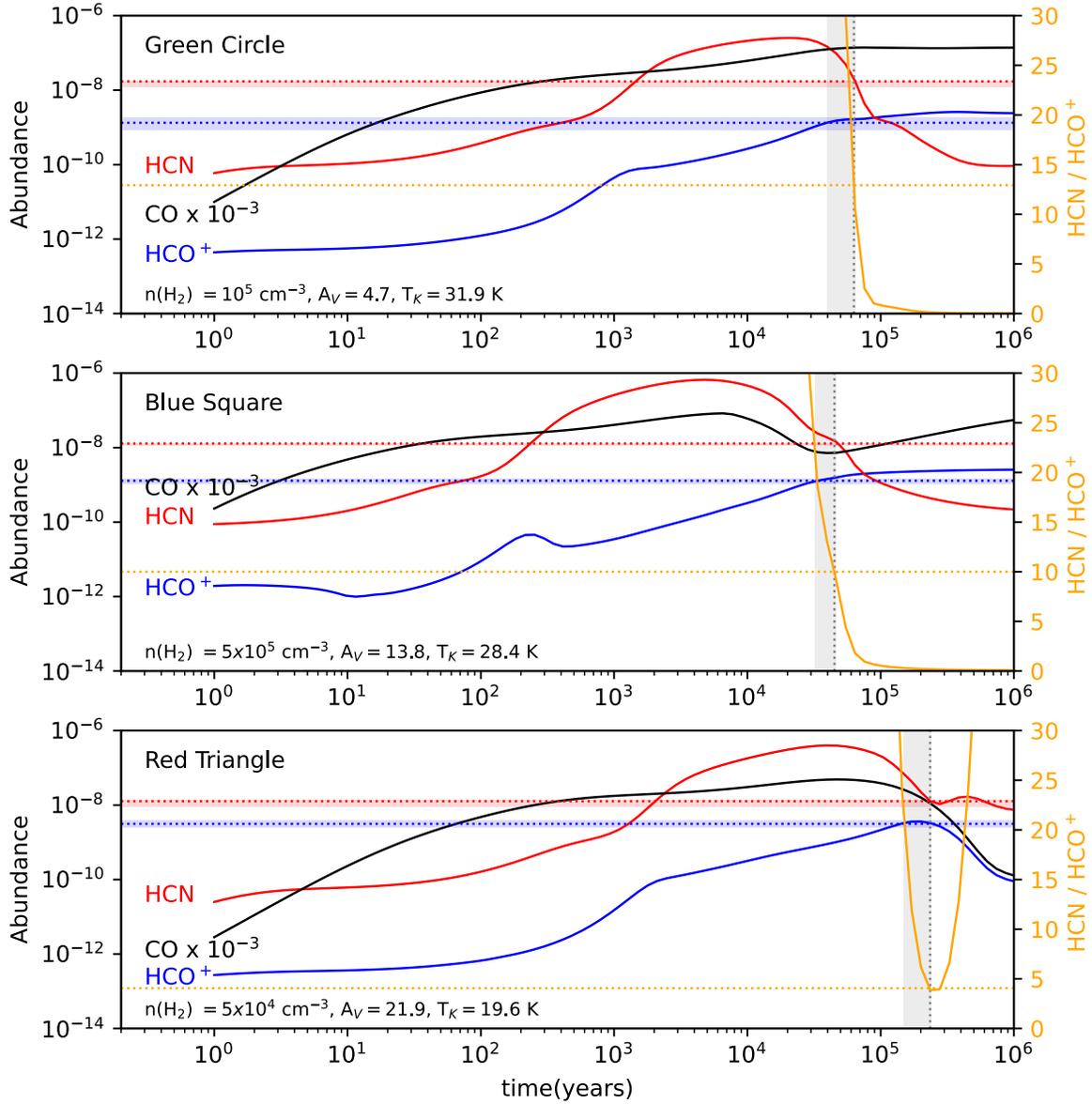

**Figure 8.** NAUTILUS model results for the three positions shown in Figure 2 (Green Circle - top panel; Blue Square - middle panel; and Red Triangle - bottom panel) showing the time evolution of HCN (red), HCO$^+$ (blue), and CO (black) abundances, and the HCN/HCO$^+$ abundance ratio (orange). For ease of comparison, the CO abundances have been scaled down by a factor of 1000 to plot them on the same scale as HCN and HCO$^+$. The horizontal dotted lines indicate the observed HCN (red dotted) and HCO$^+$ (blue dotted) adundances, and the abundance ratio (orange dotted) as determined from RADEX. The horizontal shaded regions indicate the uncertainties in the abundances. The vertical shaded region shows the time between when the RADEX and NAUTILUS HCO$^+$ abundances match and when the HCN abundances match. The vertical grey dotted line indicates the time at which the HCN/HCO$^+$ adundance ratio from the NAUTILUS model matches the observed RADEX value. The orange axis on the right shows the value of HCN/HCO$^+$ abundance ratio.

| Position | $A_V$ mag. | $T_K$ K | $n(H_2)$ cm$^{-3}$ | Model Time years | N(HCN) ($\times 10^{14}$) cm$^{-2}$ | X(HCN) ($\times 10^{-8}$) | N(HCO$^+$) ($\times 10^{13}$) cm$^{-2}$ | X(HCO$^+$) ($\times 10^{-9}$) | $\frac{X(HCN)}{X(HCO^+)}$ |
|---|---|---|---|---|---|---|---|---|---|
| Green Circle | 4.7 | 31.9 | $10^5$ | $4 - 6.5 \times 10^4$ | 1.70±0.4 | 1.72±0.5 | 1.31±0.4 | 1.33±0.5 | 12.9±6.1 |
| Blue Square | 13.8 | 28.4 | $5 \times 10^5$ | $3.2 - 4.6 \times 10^4$ | 3.75±0.3 | 1.29±0.2 | 3.75±0.3 | 1.29±0.2 | 10.0±2.2 |
| Red Triangle | 21.9 | 19.6 | $5 \times 10^4$ | $1.5 - 2.4 \times 10^5$ | 5.87±1.7 | 1.27±0.4 | 14.44±2.4 | 3.13±0.6 | 4.1±1.5 |

**Table 3.** Best model results for the HCN and HCO$^+$ abundances.

As the H$_2$ density increases, the chemical model matches the observed HCN abundances at earlier times as well. This is likely because the main formation pathway for HCN is:

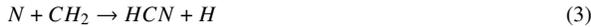

$$N + CH_2 \rightarrow HCN + H \quad (3)$$

and the main destruction pathway is:

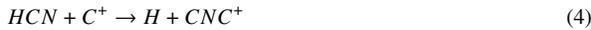

$$HCN + C^+ \rightarrow H + CNC^+ \quad (4)$$

At higher densities there is more CH$_2$ at early times to efficiently form HCN and, simultaneously, less C$^+$ to destroy HCN (the C$^+$ combines with free electrons to form neutral carbon, affecting the HCO$^+$ abundance). In comparison, in their study of CN and HCN in the Serpens cloud, Mirocha et al. (2021) found that equations 3 and 4 are not the dominant HCN reactions for a UV field strength of $G_o = 237$ in $A_V = 5$ gas. In their models, these are the dominant reactions for much weaker fields (i.e. $G_o = 10^{-3} - 10^{-1}$). The difference may lie in the fact that they used a gas temperature of 50 K in their models that can change which reactions dominate.

To investigate the effect of density on HCN and HCO$^+$ line strengths, we again used RADEX to model the expected HCN and HCO$^+$ J = 3 → 2 integrated intensities but now using the abundances and densities from Table 3. We modelled the HCN and HCO$^+$ J = 3 → 2 emission from the Red Triangle (the peak of the H$_2$ map) and Blue Square (near the peak of the HCN and HCO$^+$ emission). For the Red Triangle the model produces $T_{mb}$ (HCN J = 3 → 2) = 1.9 K and $T_{mb}$ (HCO$^+$ J = 3 → 2) = 4.4 K, which is very close to the observed values of 1.8 K and 4.5 K respectively. For the Blue Square, the model produces $T_{mb}$ (HCN J = 3 → 2) = 7.5 K and $T_{mb}$ (HCO$^+$ J = 3 → 2) = 6.9 K which is comparable to the observed values of 6.4 K and 6.8 K respectively. Thus, the reason the HCN and HCO$^+$ integrated intensity maps peak in the lower H$_2$ column density Blue Region, rather than in the high H$_2$ column density Red Region is due to excitation and abundance effects. Our RADEX results are, therefore, consistent with a lower density in the Red Region.

### 4.4 Comparison with previous work

Figure 9 shows our best-fit abundances (Table 3) alongside literature values (Table 1) for comparison. The HCN abundances in W40 (Table 3) are roughly an order of magnitude *higher* than those found in low-mass protostars and in W51, but comparable to the Young Stellar Object GL2591. They also fall within the lower range of values reported for LIRGs and ULIRGs. The enhanced HCN abundances in these extreme systems have been attributed by Nishimura et al. (2024) to chemical processing in shocks.

The HCO$^+$ abundances we derive are a factor of a few, to an order of magnitude, *lower* than those reported in most other studies, with the exception of W51 and 70 $\mu$m–quiet sources. One caveat for W51 is that Watanabe et al. (2017) adopt a uniform gas temperature of 20 K across the cloud, whereas we use dust-based temperature maps that provide $T_K$ on a pixel-by-pixel basis.

Despite variations in the absolute HCN and HCO$^+$ abundances across sources, the HCN/HCO$^+$ abundance ratios remain relatively consistent, differing by only a factor of a few between studies. AGN and ULIRG sources, however, show a broader range (e.g. Butterworth et al. 2025; Nishimura et al. 2024; Imanishi et al. 2023a; Imanishi et al. 2023b), which could be attributed to elevated HCN abundances in some systems—potentially driven by enhanced cosmic-ray ionization (Butterworth et al. 2025) or, again, by chemical processing in shocks (Nishimura et al. 2024).

Thus, our modelling of the HCN and HCO$^+$ abundances have revealed a few interesting results:

- The HCN and HCO$^+$ abundances in W40 differ from those reported in both low-mass and high-mass star-forming regions, as well as in LIRGs and ULIRGs. These discrepancies may arise from variations in physical conditions (e.g., H$_2$ density, temperature, or the presence of shocks), which can alter the rates and efficiencies of key chemical reactions.
- The HCN/HCO$^+$ abundance ratios are comparatively uniform across sources and appear to be less sensitive to differences in physical conditions.
- Abundance maps produced under the assumption of a constant density—such as those in Figure 7, which suggest variations of nearly a factor of ~ 20 across W40—may not accurately represent the true chemical structure of the region.
- By incorporating a priori knowledge of the physical environment (notably N(H$_2$) and $T_K$ from the HGBS maps), the combined use of RADEX and NAUTILUS enables us to estimate the H$_2$ volume density and derive self-consistent HCN and HCO$^+$ abundances in W40 that remain relatively uniform across the cloud.

### 4.5 Effects of the UV field and cosmic-ray ionization rate

The abundances of many species including HCN and HCO$^+$ may be affected by the strength of the UV field since, in increased UV and/or low $A_V$ environments, photochemical reactions may be enhanced. The NAUTILUS code calculates the rate coefficients for photochemical processes via the following equation (Wakelam et al. 2012):

$$k_{photo} = Ae^{-CA_V} \; s^{-1} \quad (5)$$

where A is the unattenuated photodissociation rate, which scales with the strength of the external UV field, C is a coefficient that accounts for the photodissociation/photoionization threshold of the molecular species, and $A_V$ is the visual extinction produced by the dust. Thus, the exponential factor accounts for dust attenuation of the external UV field. Increasing the external UV field will, therefore, increase the parameter A, which increases the rate coefficient.

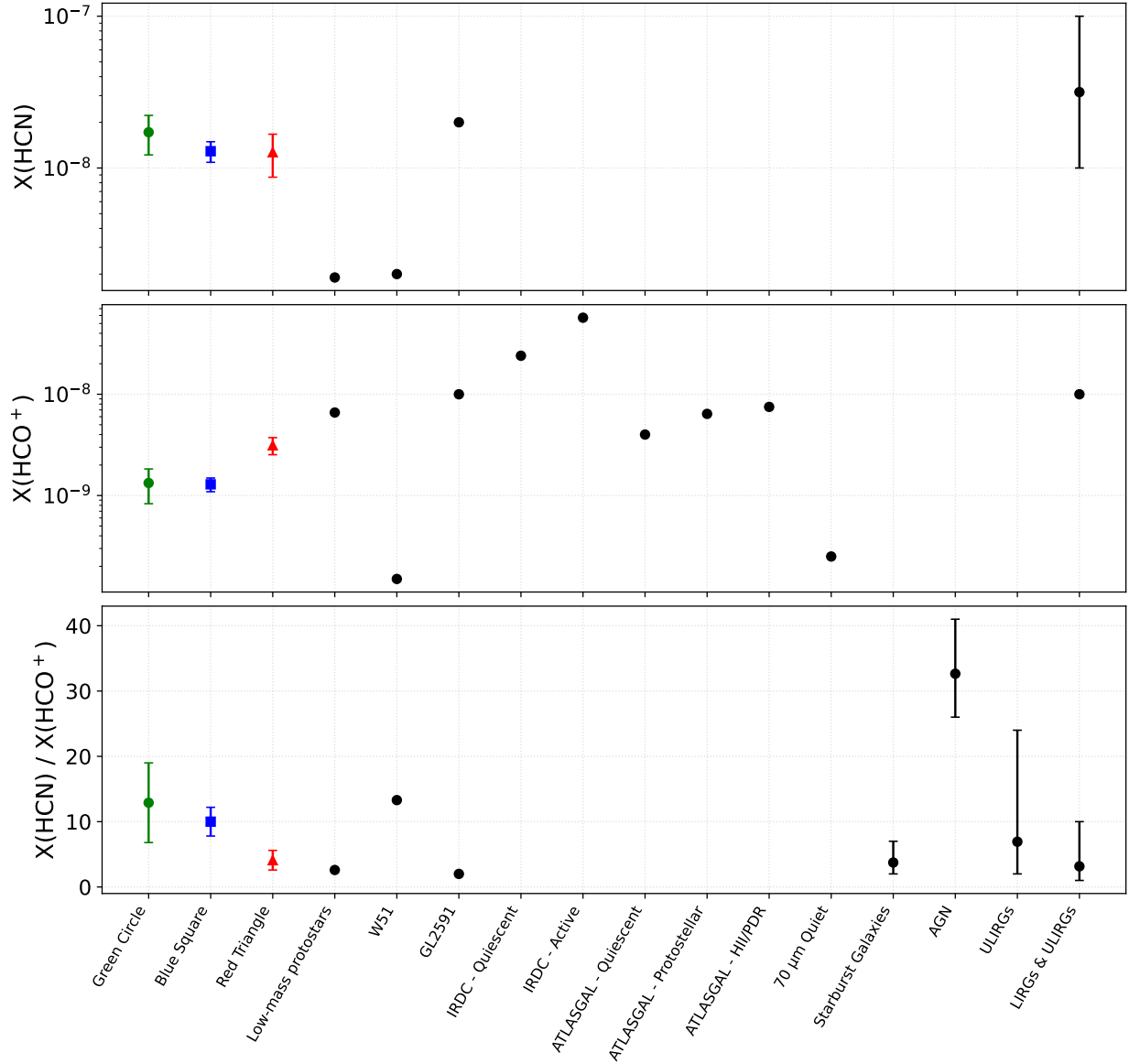

**Figure 9.** Comparison of HCN and HCO$^+$ abundances and abundance ratios in W40 with values reported for a range of Galactic and extragalactic environments. *Top:* fractional HCN abundance $X$(HCN) relative to H$_2$. *Middle:* fractional HCO$^+$ abundance $X$(HCO$^+$). *Bottom:* abundance ratio $X$(HCN)/$X$(HCO$^+$). The coloured symbols with error bars show the three W40 positions (Green Circle, Blue Square and Red Triangle), while the black symbols show literature values for low-mass protostars; W51; GL 2591; quiescent and active IRDCs; quiescent, protostellar and H<span>II</span>/PDR ATLASGAL clumps; 70 $\mu$m-quiet clumps; and integrated measurements for starburst galaxies, AGN, ULIRGs (Butterworth et al. 2025), and a combined study of LIRGs+ULIRGs (Nishimura et al. 2024). See Table 1 for complete references for each of the columns. Vertical error bars denote 1-$\sigma$ uncertainties for the W40 points and the range of values reported for each comparison sample, where available. The abundance panels (top and middle) use logarithmic y-axes, while the ratio panel (bottom) uses a linear scale.

Conversely, increasing $A_V$ will increase the attenuation of the UV field thereby decreasing the rate coefficient.

The value of $A_V$ is fixed in each pixel from the HGBS maps of N(H$_2$). The strength of the UV field is likewise fixed by the Herschel observations that show that the average strength of the UV field in W40 is 237 $G_o$ (Schneider et al. (2020); and references therein). However, the same observations show that the UV field strength can reach $\sim 8200~G_o$ near the OB association. These higher UV field strengths might be important, particularly at the position of the Green Circle (Figure 2) which is closest to the OB association, where the value of $A_V$ is lowest, and where the porous and fractal properties of molecular clouds may allow for increased UV penetration.

To test the effects of a stronger UV field on our results, we reran the NAUTILUS code for each of the three selected positions using a UV scaling factor of 8200. The results for the Blue Square and Red Triangle positions are unchanged. This is unsurprising given the large value of $A_V$ at these positions (13.3 and 21.9 respectively) that attenuates the radiation field by a factor of $1.7 \times 10^{-6}$ and $3 \times 10^{-10}$ respectively; meaning that photoprocesses are unimportant for either choice of the UV field strength. A similar conclusion was reached by Baan et al. (2010) who modelled the chemical evolution of the ISM in luminous IR galaxies and found that in high column density regions, changes in the UV radiation field only leads to small changes in the HCN/ HCO$^+$ line ratios, because the UV flux is largely attenuated.

At the position of the Green Circle, a radiation field of 8200 $G_o$, however, is unable to produce the observed HCN and HCO$^+$ abundances. In particular, over the timespan of the simulation, the model produces less HCN (up to a factor of $\sim$ 80) and HCO$^+$ (up to a factor of $\sim$ 20) than is observed. This is likely because the value of $A_V$ is only 4.7, which provides an attenuation of only $9 \times 10^{-3}$; meaning that photoprocesses are likely important. The main processes lowering the abundances of HCN and HCO$^+$ are not direct photodissociations as might be expected but, instead, are Equation 4 and:

$$HCN + H^+ \rightarrow H + HNC^+ \qquad (6)$$

for HCN, and:

$$HCO^+ + e^- \rightarrow H + CO \qquad (7)$$

for HCO$^+$. The magnitude of the UV field, coupled with the weak attenuation, means that photoprocesses produce an ample amount of C$^+$, H$^+$, and e$^-$ to efficiently destroy both species through gas-phase reactions.

These results are slightly different from those found by Mirocha et al. (2021) using $G_o = 1 - 10^6$, $T_K = 50$ K, and $A_V = 5$, in which the dominant destruction reaction is:

$$HCN + h\nu \rightarrow H + CN \qquad (8)$$

instead of equations 4 and 6. In their models, equations 4 and 6 are the dominant destruction reactions for much weaker fields (i.e. $G_o = 10^{-3} - 10^{-1}$). Again the difference may be due to the higher temperatures that they use in their simulations.

We also explored the value of $A_V$ required to significantly reduce the impact of the UV field and, by $A_V = 6$, the HCN abundance is decreased by, at most, a factor of $\sim$ 3 and HCO$^+$ by $\sim$ 1.5. Note, however, that only 54 pixels (16 per cent) have values of $A_V \leq 6$. Of these, 42 lie to the left and below of the Green Region, 12 lie within the Green Region (including the pixel corresponding to the Green Circle), and none lie in the Blue or Red Regions. Thus, given the high value of $A_V$ across most of W40, enhanced UV field strengths are probably irrelevant to our overall conclusions.

In order to match the observed abundances and the abundance ratio at the position of the Green Circle using a UV field of 8200 $G_o$ and the observed $A_V$ of 4.7, we have to increase the H$_2$ density by a factor of 10. For a density of n(H$_2$) = $10^6$ cm$^{-3}$ the RADEX derived abundances are $X$(HCN) = $1.3 \times 10^{-9}$ and $X$(HCO$^+$) = $2.0 \times 10^{-10}$, which are 13.2 times and 6.7 times (respectively) lower than the values in Table 3 for n(H$_2$) = $10^5$ cm$^{-3}$, but, oddly, similar to the abundances found by Traficante et al. (2017) in 70$\mu$m **quiet** sources, and in W51 (Watanabe et al. 2017) (albeit for a different H$_2$ density). The fact that the HCN abundance decreases more than that of HCO$^+$ is probably due to the fact that, for the first $10^5$ years, the HCN destruction rate from equation 4 is an order of magnitude larger than that for HCO$^+$ (equation 7). The HCN/ HCO$^+$ abundance ratio, therefore, is decreased by a factor $\sim$ 2 from 12.9 to 6.7. In this NAUTILUS model, both observed abundances and the abundance ratio are reached at $\sim$ $10^4$ years.

The HCO$^+$ abundance is also sensitive to the cosmic-ray (CR) ionization rate, since its primary formation route is via H$_3^+$ (see Equation 1) which is itself strongly dependent on the cosmic-ray flux. The two steps leading up to Equation 1 are:

$$H_2 + CR \rightarrow H_2^+ + e^- \qquad (9)$$

and

$$H_2^+ + H_2 \rightarrow H_3^+ + H \qquad (10)$$

.

In our models we adopt a "standard" cosmic-ray ionization rate of $\zeta_{H_2} = 1.3 \times 10^{-17}$ s$^{-1}$, which is appropriate for dense molecular clouds where attenuation reduces the CR flux (Caselli et al. 1998; Padovani et al. 2009). However, observations along several lines of sight toward massive star-forming regions (Indriolo et al. 2015) indicate a higher mean value of $\zeta_{H_2} = 1.78 \times 10^{-16}$ s$^{-1}$. To test the impact of an enhanced CR ionization rate, we ran additional models using the Indriolo et al. (2015) value together with our fiducial UV field strength (237 $G_o$).

While there are differences in the time evolution of each species (and their ratio) between the two models, the principal effect of increasing $\zeta_{H_2}$ to $1.78 \times 10^{-16}$ s$^{-1}$ is that the steady-state abundances of both HCN and HCO$^+$ (at times $\gtrsim 10^5$ yr) are higher. However, with the elevated ionization rate, the Red Triangle and Green Circle density solutions remain unchanged (i.e. $5 \times 10^5$ and $10^5$ cm$^{-3}$ respectively). The Blue Square, however, is best reproduced by a NAUTILUS model with a density of $10^5$ cm$^{-3}$ rather than $5 \times 10^5$ cm$^{-3}$. See Appendix B for further details on the model results using the enhanced cosmic-ray ionization rate.

From this analysis, therefore, we can draw some additional conclusions:

- In the presence of a strong UV field and **high** $A_V$ the abundances of HCN and HCO$^+$ are *not* affected due to attenuation of the radiation field in regions of high H$_2$ column density.
- In the presence of a strong UV field and **low** $A_V$ the abundance of both HCN and HCO$^+$ are *strongly* affected. The HCN abundance, however, is decreased by a factor of $\sim$ 2 more than the HCO$^+$ abundance.
- Increasing the cosmic-ray ionization rate changes the detailed time evolution of HCN and HCO$^+$ and produces higher steady-state abundances, but it does not strongly affect the density solutions of our models.

## 5 SUMMARY AND CONCLUSIONS

As part of a pilot project for the Massive, Active, JCMT-Observed Regions of Star Formation (MAJORS) Large Program at the JCMT, we mapped the HCN and HCO$^+$ $J = 3 \rightarrow 2$ transitions across the central 424″ × 424″ region of the massive star-forming complex W40. Using results from the Herschel Gould Belt Survey (HGBS), we assigned a kinetic temperature (assuming $T_K = T_{dust}$) and an H$_2$ column density $N$(H$_2$) to every pixel in our HCN and HCO$^+$ maps. With $T_K$ and the integrated intensities of HCN and HCO$^+$ as inputs to the RADEX 1D radiative transfer code, we computed the corresponding HCN and HCO$^+$ column densities for a range of H$_2$ densities ($10^4$–$10^6$ cm$^{-3}$). Abundances in each pixel were then obtained by dividing these column densities by the HGBS-derived values of $N$(H$_2$). Because the RADEX-derived column densities and abundances depend on the assumed H$_2$ density, we subsequently used the NAUTILUS gas–grain chemical evolution code to determine the H$_2$ density at which the NAUTILUS abundances, and the HCN/HCO$^+$ abundance ratios, match those obtained from RADEX.

The main results from our study are as follows:

- The bulk of the HCN and HCO$^+$ emission in W40 is not co-located with the region of maximum H$_2$ column density but, instead, arises in the Blue Region (see Figure 2) which, from our modelling, contains the highest H$_2$ volume density and intermediate dust temperatures (i.e. 26-28 K). In addition, the ratio of the HCN to HCO$^+$ integrated intensities, $I$(HCN)/$I$(HCO$^+$), increases with H$_2$ density. However, care must be taken in assigning a specific line intensity and $I$(HCN)/$I$(HCO$^+$) ratio to a specific H$_2$ density, since different excitation conditions and molecular abundances will produce different intensities and ratios at the same density.

- By incorporating a priori knowledge of the physical conditions (notably the N(H$_2$) and $T_K$ data from the HGBS maps) and assuming a constant density of n(H$_2$) = $10^5$ cm$^{-3}$ we used RADEX to calculate the HCN and HCO$^+$ abundances in every pixel in our W40 map. The HCN abundances range from $0.4 - 7.0 \times 10^{-8}$ and the HCO$^+$ adundances range from $0.4 - 7.3 \times 10^{-9}$. The corresponding HCN/HCO$^+$ abundance ratio range is $2.6 - 23.1$.

- Using the NAUTILUS chemical code to find the H$_2$ density that produces the best match to the RADEX abundances (rather than a constant density) suggests, instead, that the HCN and HCO$^+$ abundances may be fairly constant across W40. Careful modelling of three different positions (with disparate physical conditions) finds $X$(HCN) = $1.3 - 1.7 \times 10^{-8}$, $X$(HCO$^+$) = $1.3 - 3.1 \times 10^{-9}$. The HCN/HCO$^+$ abundance ratio ranges from $4.1 - 12.9$.

- By cross-comparing the results of RADEX and NAUTILUS, we obtain a crude estimate of the gas density responsible for the HCN and HCO$^+$ emission, finding H$_2$ densities in the range $5 \times 10^4 - 5 \times 10^5$ cm$^{-3}$.

- High UV intensity has little effect on the HCN and HCO$^+$ abundance in regions where the visual extinction is large enough to effectively shield the gas from the UV field. In regions where $A_V$ is low (i.e. < 6), however, the abundance of both species is lowered due to destructive reactions with species which are directly affected by the radiation field (e.g. H$^+$, C$^+$, and e$^-$). In such regions, our models indicate that the HCN abundance decreases by a factor of $\sim$ 13 whereas the HCO$^+$ abundance decreases by a factor of 6. Thus, decreased HCN and HCO$^+$ abundances, coupled with decreased HCN/HCO$^+$ abundance ratios, may indicate the presence of a strong UV field (and low extinction). Care must be taken with this interpretation, however, since variations in these quantities can also be a result of varying physical conditions (e.g. n(H$_2$) and $T_K$) in a weak UV field.

**ACKNOWLEDGEMENTS**

The James Clerk Maxwell Telescope is operated by the East Asian Observatory on behalf of The National Astronomical Observatory of Japan; Academia Sinica Institute of Astronomy and Astrophysics; the Korea Astronomy and Space Science Institute; centre for Astronomical Mega-Science (as well as the National Key R&D Program of China with No. 2017YFA0402700). Additional funding support is provided by the Science and Technology Facilities Council of the United Kingdom and participating universities in the United Kingdom and Canada. The James Clerk Maxwell Telescope has historically been operated by the Joint Astronomy Centre on behalf of the Science and Technology Facilities Council of the United Kingdom, the National Research Council of Canada and the Netherlands Organisation for Scientific Research.

This research has made use of data from the Herschel Gould Belt survey (HGBS) project (http://gouldbelt-herschel.cea.fr). The HGBS is a Herschel Key Programme jointly carried out by SPIRE Specialist Astronomy Group 3 (SAG 3), scientists of several institutes in the PACS Consortium (CEA Saclay, INAF-IFSI Rome and INAF-Arcetri, KU Leuven, MPIA Heidelberg), and scientists of the Herschel Science centre (HSC). This publication also makes use of data products from the Two Micron All Sky Survey, which is a joint project of the University of Massachusetts and the Infrared Processing and Analysis centre/California Institute of Technology, funded by the National Aeronautics and Space Administration and the National Science Foundation.

RP acknowledges the support of the Natural Sciences and Engineering Research Council of Canada (NSERC), through the Discovery Grant program and JC's work was supported by the NSERC USRA program. Researchers at the University of Calgary also acknowledge and pay tribute to the traditional territories of the peoples of Treaty 7, which include the Blackfoot Confederacy (comprised of the Siksika, the Piikani, and the Kainai First Nations), the Tsuut'ina First Nation, and the Stoney Nakoda (including Chiniki, Bearspaw, and Goodstoney First Nations). The City of Calgary is also home to the Métis Nation of Alberta Region 3. The authors also wish to recognize and acknowledge the very significant cultural role and reverence that the summit of Maunakea has always had within the indigenous Hawaiian community. We are most fortunate to have the opportunity to conduct observations from this mountain.

LCH was supported by the China Manned Space Program (CMS-CSST-2025-A09), the National Science Foundation of China (12233001), and the National Key R&D Program of China (2022YFF0503401) DA acknowledges NU FCDRP grant 201223FDD8821. PS was partially supported by a Grant-in-Aid for Scientific Research (KAKENHI Number JP22H01271 and JP23H01221) of JSPS. PS was supported by Yoshinori Ohsumi Fund (Yoshinori Ohsumi Award for Fundamental Research). K.P. is a Royal Society University Research Fellow, supported by grant number URF\R1\211322. SER acknowledge support from consolidated grant (ST\W000830\1) from the UK Science and Technology Facilities Council (STFC). The work of MGR is supported by the international Gemini Observatory, a program of NSF NOIRLab, which is managed by the Association of Universities for Research in Astronomy (AURA) under a cooperative agreement with the U.S. National Science Foundation, on behalf of the Gemini partnership of Argentina, Brazil, Canada, Chile, the Republic of Korea, and the United States of America. C.W.L is supported by the Basic Science Research Program through the NRF funded by the Ministry of Education, Science and Technology (grant No. NRF-2019R1A2C1010851) and by the Korea Astronomy and Space Science Institute grant funded by the Korea government (MSIT; project No. 2024-1-841-00). MGB acknowledges the support of the Department for Communities in Northern Ireland for support of Armagh's contribution to the JCMT and to this research programme. AK acknowledges support from the Polish National Science Center grant No. 2024/54/E/ST9/00314.

## 6 DATA AVAILABILITY

The MAJORS survey and its data products are described fully in Eden et al. (2026). The associated download instructions are within that paper, or per request to the lead author. Observational data from the JCMT are publicly available through the Canadian Astron-omy Data Centre (CADC) at https://www.cadc-ccda.hia-iha.nrc-cnrc.gc.ca/en/ under the proposal IDs M22AL002 and M20AD003.

This paper has been typeset from a TEX/LATEX file prepared by the author.

## APPENDIX A: MODEL FITTING DETAILS

In this appendix, we present the model results for each of our three positions at three different densities and demonstrate how the best-matching solution is selected. Descriptions are provided in the figure captions.

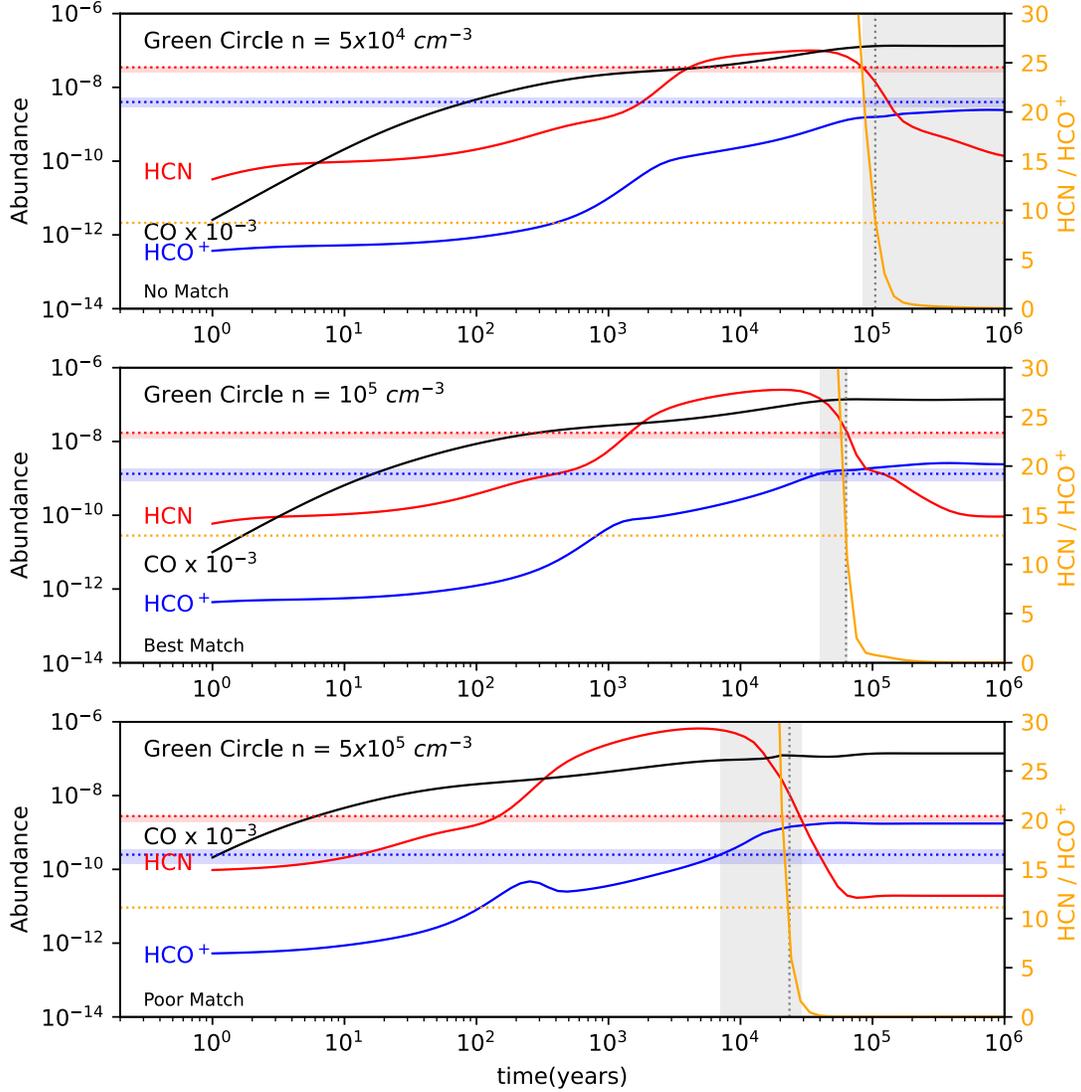

Figure A1: NAUTILUS model results for the Green Circle showing the time evolution of HCN (red), HCO$^+$ (blue), and CO (black) abundances, and the HCN/HCO$^+$ abundance ratio (orange) for three different H$_2$ densities – $5 \times 10^4$ (top), $10^5$ (middle), and $5 \times 10^5$ (bottom) cm$^{-3}$. This figure follows the same conventions as Figure 8, with identical meanings for the vertical and horizontal dashed lines as well as the shaded regions. This figure shows that an H$_2$ density of $10^5$ cm$^{-3}$ (middle) provides the best match to the observed abundances. At a density of $5 \times 10^4$ cm$^{-3}$ (top), the NAUTILUS model fails to reproduce the observed HCO$^+$ abundance at any time. While the grey shaded regions for densities of $10^5$ cm$^{-3}$ (middle) and $5 \times 10^5$ cm$^{-3}$ (bottom) span similar absolute time intervals (∼ $2.5 \times 10^4$ years), we select the middle panel as the best match because the HCN crossing time is only 1.6 times larger than the HCO$^+$ crossing time, compared with 4.1 times larger in the bottom panel.

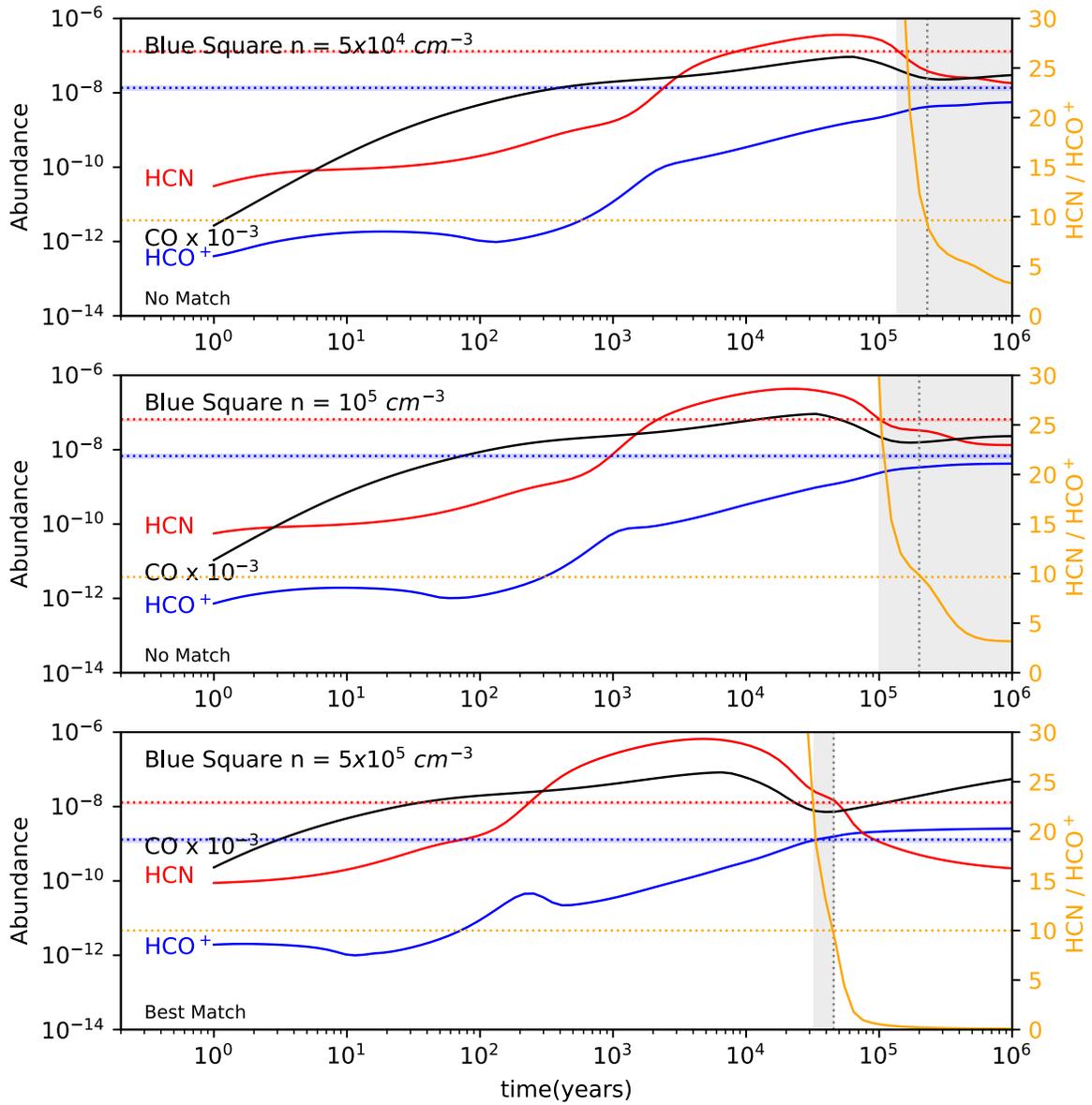

**Figure A2.** Same as Figure A1 except that results are shown for the Blue Square. This figure clearly shows that a density of $5 \times 10^5$ cm$^{-3}$ (bottom) provides the best match to the abundances since, at the other two densities, the NAUTILUS model fails to reproduce the observed HCO$^+$ abundances at any time.

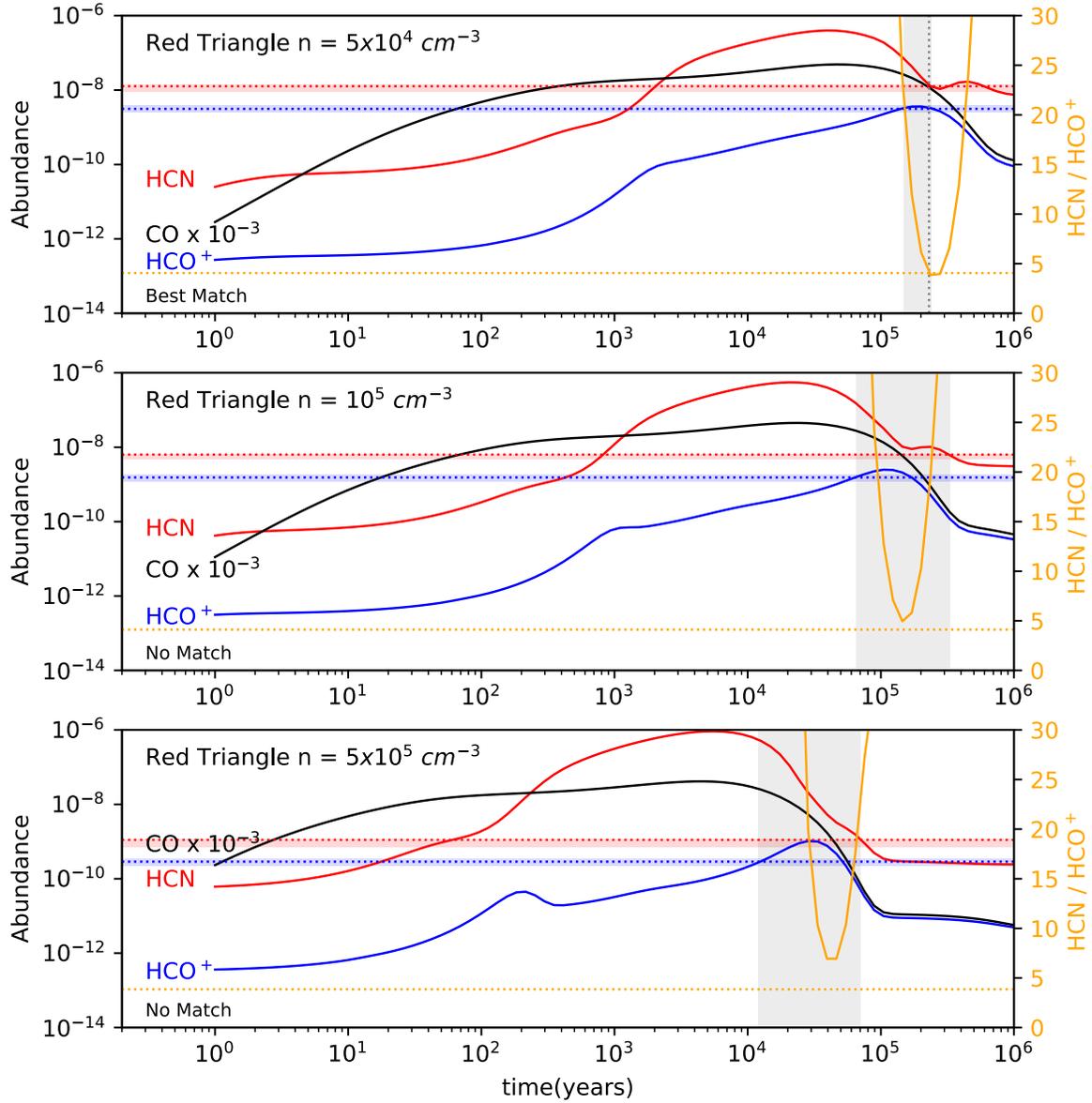

**Figure A3.** Same as Figure A1 except that results are shown for the Red Triangle. This figure clearly shows that a density of $5 \times 10^4$ cm$^{-3}$ (top) provides the best match to the abundances since, at the other two densities, the NAUTILUS model fails to reproduce the observed HCN/HCO$^+$ abundance *ratios* at any time.

## APPENDIX B: ENHANCED COSMIC-RAY IONIZATION RATE MODELS

In this appendix, we present the model results for each of our three positions at three different densities using an enhanced cosmic-ray ionization rate of $\zeta_{H_2} = 1.78 \times 10^{-16}$ s$^{-1}$). Descriptions are provided in the figure captions.

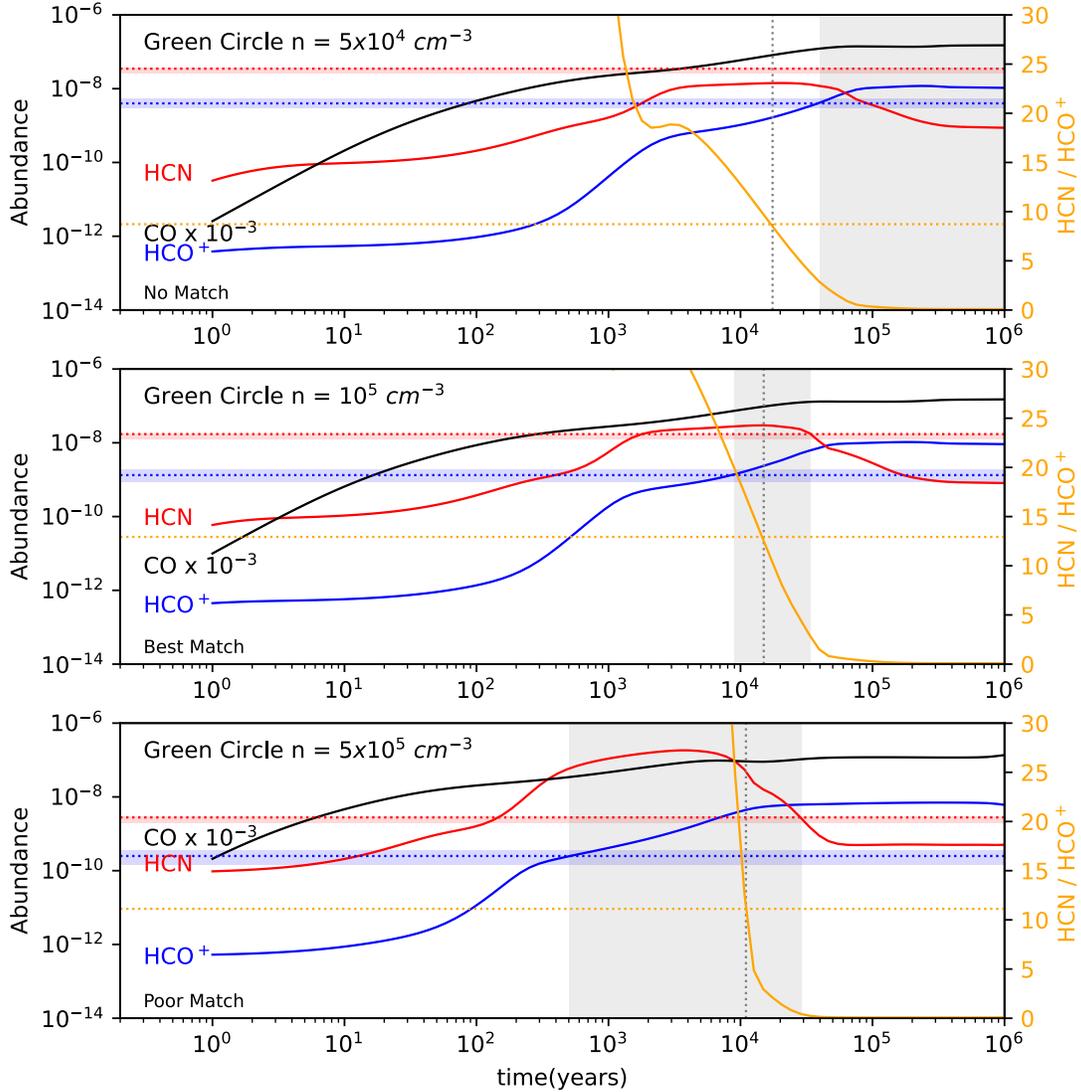

Figure B1: NAUTILUS model results (using an enhanced cosmic-ray ionization rate of $\zeta_{H_2} = 1.78 \times 10^{-16}$ s$^{-1}$) for the Green Circle showing the time evolution of HCN (red), HCO$^+$ (blue), and CO (black) abundances, and the HCN/HCO$^+$ abundance ratio (orange) for three different H$_2$ densities - $5 \times 10^4$ (top), $10^5$ (middle), and $5 \times 10^5$ (bottom) cm$^{-3}$. This figure follows the same conventions as Figure 8, with identical meanings for the vertical and horizontal dashed lines as well as the shaded regions. This figure shows that an H$_2$ density of $10^5$ cm$^{-3}$ (middle) provides the best match to the observed abundances. At a density of $5 \times 10^4$ cm$^{-3}$ (top), the NAUTILUS model fails to reproduce the observed HCN abundance at any time, while the grey shaded region for a density of $5 \times 10^5$ cm$^{-3}$ (bottom) spans a larger time interval than that for $10^5$ cm$^{-3}$ (middle). This density solution is the same as found for the "standard" cosmic-ray ionization rate of $\zeta_{H_2} = 1.3 \times 10^{-17}$ s$^{-1}$ (see Figure 8).

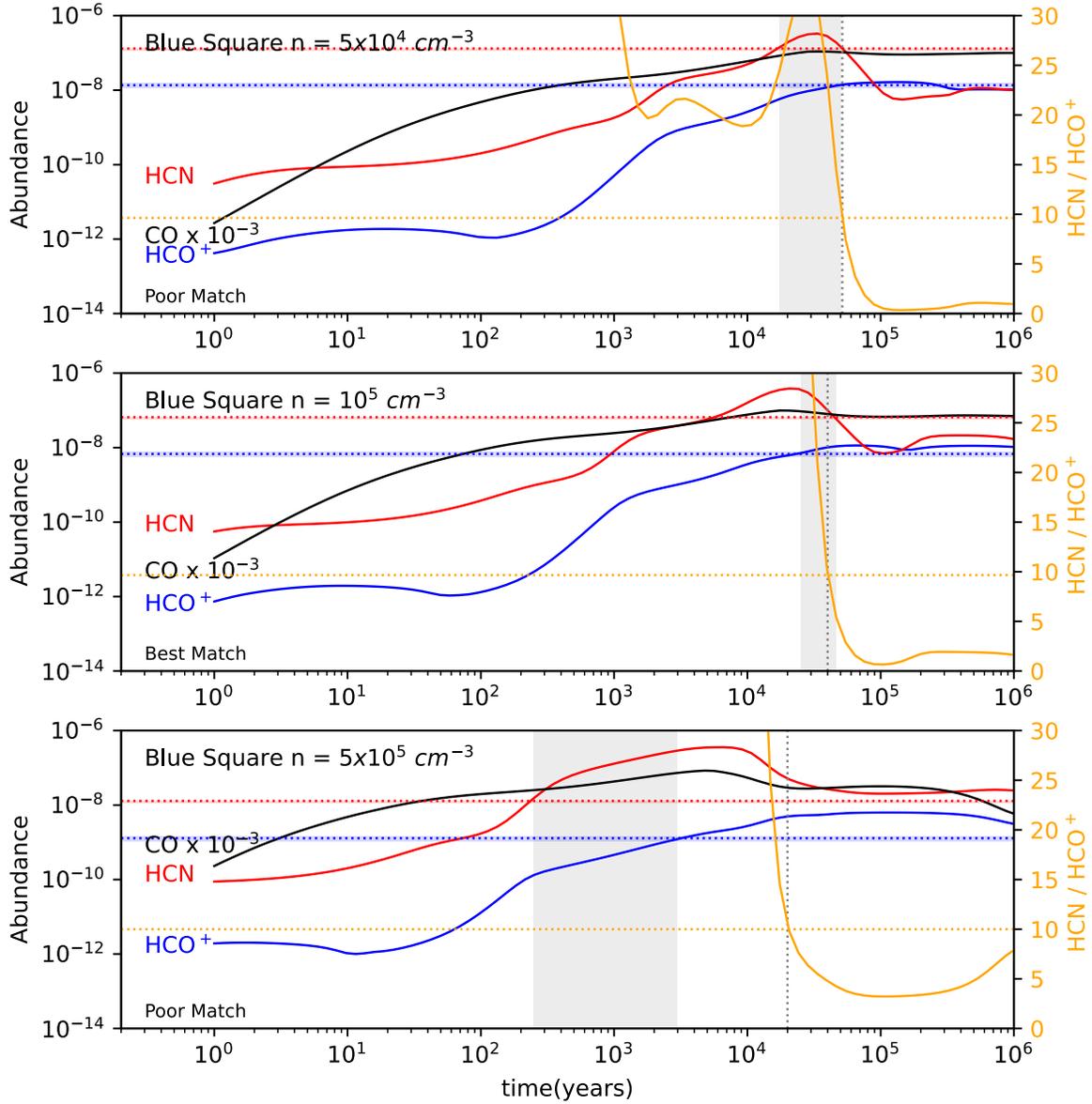

**Figure B2.** Same as Figure B1 except that results are shown for the Blue Square. This figure clearly shows that a density of $10^5$ cm$^{-3}$ (middle) provides the best match to the abundances. The top panel ($5 \times 10^4$ cm$^{-3}$) exhibits the largest time difference between the HCN and HCO$^+$ model crossing times. The bottom panel ($5 \times 10^5$ cm$^{-3}$) has the shortest time difference, but the abundance ratio falls outside this overlap window. Notably, this preferred density differs from the solution found for the "standard" cosmic-ray ionization rate of $\zeta_{H_2} = 1.3 \times 10^{-17}$ s$^{-1}$, which favored $5 \times 10^5$ cm$^{-3}$ (see Figure 8).

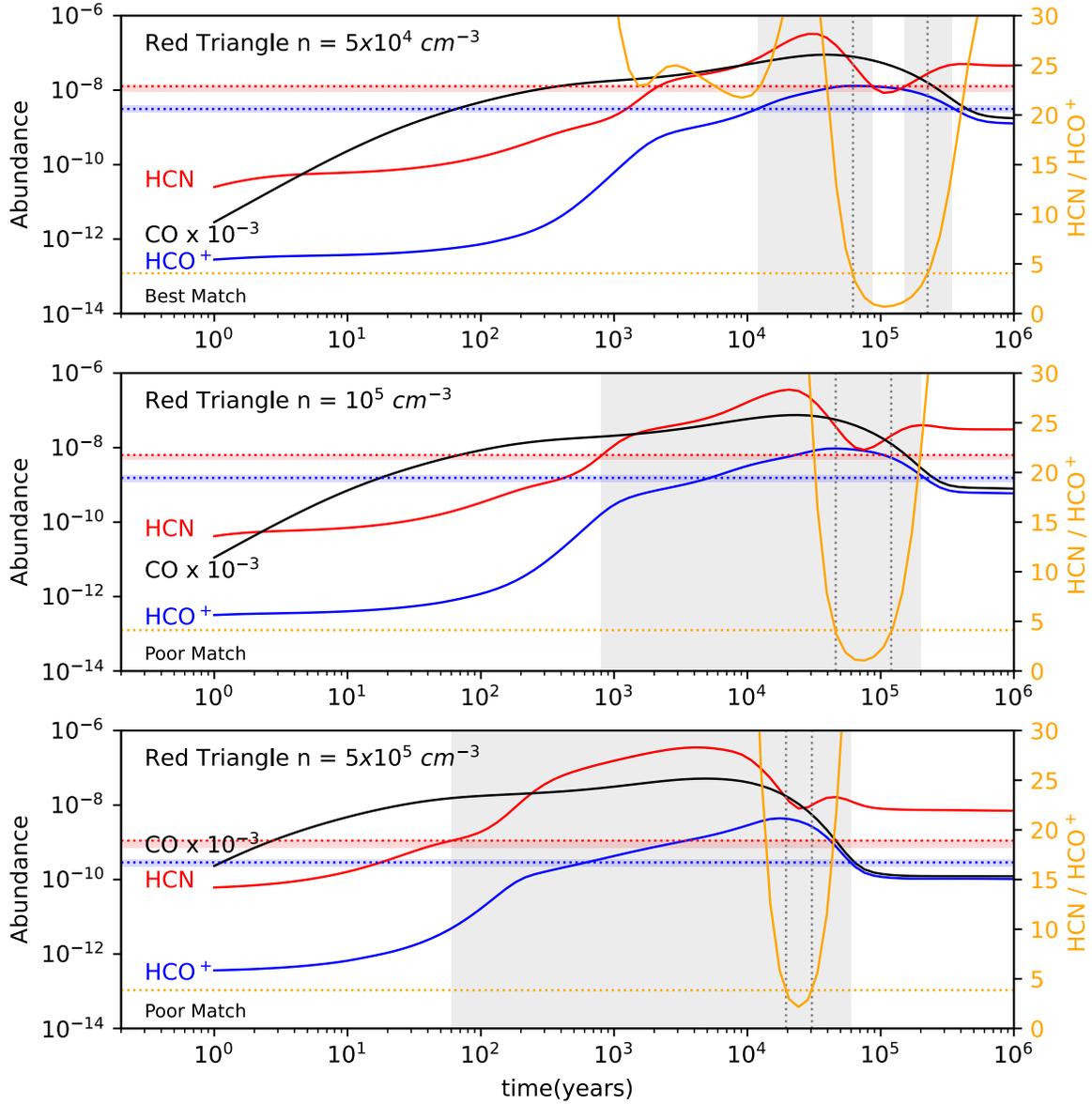

**Figure B3.** Same as Figure B1 except that results are shown for the Red Triangle. This Figure shows that there are two times at which the HCN/HCO⁺ abundance ratio matches the observed value. In the bottom panel, both ratio solutions fall within a broad time interval over which the HCN and HCO⁺ models cross the observed values—spanning nearly three orders of magnitude in time. In the middle panel, the difference between the HCN and HCO⁺ model crossing times exceeds two orders of magnitude. In the top panel, there are two distinct possible solutions: one between $1.2-8.4 \times 10^4$ years and another between $1.5-3.4 \times 10^5$ years, each spanning a factor of a few in time. Thus, as in the case of the "standard" cosmic-ray ionization rate (Figure 8), the best-fitting density is $5 \times 10^4$ cm$^{-3}$ (top panel).